\documentclass[fleqn,usenatbib]{mnras}

\usepackage{newtxtext,newtxmath}

\usepackage[T1]{fontenc}

\DeclareRobustCommand{\VAN}[3]{#2}
\let\VANthebibliography\thebibliography
\def\thebibliography{\DeclareRobustCommand{\VAN}[3]{##3}\VANthebibliography}

\usepackage{graphicx}	
\usepackage{amsmath}	
\usepackage{ulem}
\defcitealias{Lanzuisi_2018}{L18}

\title[Obscured AGN demographics]{The demographics of obscured AGN from X-ray spectroscopy guided by multiwavelength information}

\author[Brivael Laloux]{
Brivael Laloux$^{1,2}$\thanks{E-mail: brivael.laloux@noa.gr}, 
Antonis Georgakakis$^1$,
Carolina Andonie$^2$,
David M. Alexander$^2$,
Angel Ruiz$^1$, \newauthor
David J. Rosario$^{2,3}$,
James Aird$^{4,5}$,
Johannes Buchner$^6$,
Francisco J. Carrera$^7$,
Andrea Lapi$^8$, \newauthor
Cristina Ramos Almeida$^{9.10}$,
Mara Salvato$^{6,11}$, 
Francesco Shankar$^{12}$
\\
$^{1}$ Institute for Astronomy \& Astrophysics, National Observatory of Athens, V. Paulou \& I. Metaxa, 11532, Greece\\
$^{2}$ Centre for Extragalactic Astronomy, Department of Physics, Durham University, UK\\
$^{3}$ School of Mathematics, Statistics and Physics, Newcastle University, Newcastle upon Tyne, NE1 7RU, UK\\
$^{4}$ Institute for astronomy, University of Edinburgh, Royal observatory, Edinburgh, EH9 3HJ, UK\\
$^{5}$ School of Physics \& Astronomy, University of Leicester, University road, Leicester, LE1 7RJ, UK\\
$^6$ Max Planck Institute for Extraterrestrial Physics, Giessenbachstrasse, 85741 Garching, Germany\\
$^7$ Instituto de Física de Cantabria (CSIC - Universidad de Cantabria), Avenida de los Castros, 39005 Santander, Spain\\
$^8$ SISSA, Via Bonomea 265, 34136 Trieste, Italy\\
$^{9}$ Instituto de Astrofísica de Canarias, Calle Vía Láctea, s/n, E-38205, La Laguna, Tenerife, Spain\\
$^{10}$ Departamento de Astrofísica, Universidad de La Laguna, E-38206 La Laguna, Tenerife, Spain\\
$^{11}$ Exzellenzcluster ORIGINS, Boltzmannstr. 2, D-85748 Garching, Germany\\
$^{12}$ Department of Physics and Astronomy, University of Southampton, Highfield, SO17 1BJ, UK\\
}

\date{Accepted XXX. Received YYY; in original form ZZZ}

\pubyear{2021}

\begin{document}
\label{firstpage}
\pagerange{\pageref{firstpage}--\pageref{lastpage}}
\maketitle

\begin{abstract}
{A complete census of Active Galactic Nuclei (AGN) is a prerequisite for understanding the growth of supermassive black holes across cosmic time. A significant challenge toward this goal is the whereabouts of heavily obscured AGN that remain uncertain. This paper sets new constraints on the demographics of this population by developing a methodology that combines X-ray spectral information with priors derived from multiwavelength observations. We select X-ray AGN in the Chandra COSMOS Legacy survey and fit their $2.2-500\mu m$ spectral energy distributions with galaxy and AGN templates to determine the mid-infrared ($6\mu m$) luminosity of the AGN component. Empirical correlations between X-ray and $6\mu m$ luminosities are then adopted to infer the intrinsic accretion luminosity at X-rays for individual AGN. This is used as prior information in our Bayesian X-ray spectral analysis to estimate physical properties, such as line-of-sight obscuration. Our approach breaks the degeneracies between accretion luminosity and obscuration that affect X-ray spectral analysis, particularly for the most heavily obscured (Compton-Thick) AGN with low photon counts X-ray spectra. The X-ray spectral results are then combined with the selection function of the Chandra COSMOS Legacy survey to derive the AGN space density and a Compton-Thick fraction of $21.0^{+16.1}_{-9.9}\%$ at redshifts $z<0.5$. At higher redshift, our analysis suggests upper limits to the Compton-Thick AGN fraction of $\la 40\%$. These estimates are at the low end of the range of values determined in the literature and underline the importance of multiwavelength approaches for tackling the challenge of heavily obscured AGN demographics.}\
\end{abstract}




\section{Introduction}

SuperMassive Black Holes (SMBHs) are found to be ubiquitous in the nuclear regions of local galaxies \citep{Kormendy2013}. These compact objects are thought to grow their masses either via accretion of material from their surroundings \citep[e.g.][]{Soltan1982, Alexander_2012} or through merging with other black holes \citep[e.g.][]{Volonteri2003, ONeill2022}. During such active periods, large amounts of energy can be produced and observed as electromagnetic radiation at different parts of the spectrum. The class of astrophysical sources that correspond to such events are broadly dubbed Active Galactic Nuclei \citep[AGN,][]{Antonucci_1993, Urry_1995, Padovani2017}. Observational measurements of the space density of AGN in the Universe as a function of cosmic time provide essential constraints on the growth history of SMBHs we observe in the local Universe \citep{Marconi2004}. Although simple in principle, counting AGN in a cosmological volume is challenging because of both observational limitations and the phenomenological complexity of active black holes. For example, the amount of energy radiated by individual accretion events brackets many orders of magnitude and strongly depends on wavelength. Therefore, accounting for lower luminosity events in flux-limited samples is not straightforward and requires a good understanding of observational biases and selection effects. Moreover, a substantial fraction of SMBHs in the Universe is believed to grow their masses behind clouds of dust and gas \citep{Maiolino1998, Risaliti_1999} that attenuate the emitted radiation and render the identification of such systems difficult. This introduces biases in AGN counting experiments and can lead to a significant underestimation of the true size of the underlying population. Accounting for this effect requires a handle on the obscuration distribution of AGN. This has motivated observational programs that aim to constrain the fraction of obscured AGN in the Universe and provide an unbiased census of the active SMBH population \cite[e.g.][]{Hickox2018}. 

Among the different wavebands available for studying AGN obscuration, the X-ray regime offers several advantages. X-ray photons, particularly at harder rest-frame energies ($\ga2$\,keV), are less affected by intervening gas clouds compared to e.g. UV/optical, and can, therefore, provide the least biased samples for demographical investigations. Besides, at these energies, the contamination of the host galaxy by X-ray binaries or supernova remnants is low. Moreover, any obscuring material along the line-of-sight (LOS) imprints characteristic signatures on the X-ray spectra of AGN. This, in turn, translates into direct measurements of the density of the intervening obscuring clouds for individual AGN since X-ray imaging observations typically also provide spectral information. Having a handle on the level of obscuration means the possibility to quantify the selection function of X-ray surveys, i.e. the probability of detecting AGN of a given intrinsic luminosity, redshift and obscuration. This key feature enables the crucial step from an observational census of AGN  (i.e. a sample) to the demographics of the underlying population. As a result, the most detailed description of the whereabouts of obscured AGN to date has been painted by high-energy survey programs \cite[e.g][]{Ueda_2003, Barger2003, DellaCeca2008, Ueda_2014, Burlon_2011, Alexander2013, Buchner_2015, Aird_2015}. 

X-ray surveys have also identified a deeply buried AGN population, for which the surrounding obscuring material is optically thick even to X-ray photons \cite[e.g.][]{Ricci_2015}. The LOS obscuration of these sources, parameterised by the neutral hydrogen column density, $N_{\rm H}$, exceeds the Thomson scattering limit, $N_{\rm H}>\rm 1.5\times10^{24}\,cm^{-2}$. These sources are often referred to as Compton-Thick (CTK). Conversely, sources with a lower column density are referred to as Compton-Thin (CTN). The direct X-ray emission of CTK sources is largely suppressed by both photoelectric absorption and Compton scattering. Their X-ray spectra are thought to be dominated by indirect radiation, i.e. photons that have been scattered off obscuring material into the LOS. This produces a characteristic spectral shape that includes a flat continuum with superimposed strong emission lines, the most prominent of which is the Fe\,K$\alpha$ at 6.4\,keV \citep[e.g.][]{Levenson2006, Nandra2007}, and an excess of high energy photons ($>10$keV) forming the so-called Compton hump \cite{Piro_1990}. However, because of the high level of obscuration, CTK AGN appear X-ray faint and are typically detected at the flux limits of current extragalactic X-ray surveys. As a result, their spectra typically suffer from low count statistics, which translates into significant uncertainties in the determination of their intrinsic properties, such as accretion luminosity and column density \citep[][]{Buchner_2015, Saha2022}. Additional information on these sources comes from the shape and normalisation of the Cosmic diffuse X-ray Background (CXB) spectrum, which is dominated by the integrated emission of all AGN throughout the Universe. The reconstruction of the CXB using AGN population synthesis models points to a potentially significant population of CTK sources, larger than the observed one \citep{Gilli2007, Akylas_2012, Ananna2019}. Nevertheless, the exact space density of this population depends on the modelling details such as the shape of the intrinsic spectrum of individual AGN \citep{Akylas_2016}.

The InfraRed (IR) part of the electromagnetic spectrum provides an alternative wavelength regime for studying heavily obscured AGN. 
Indeed, UV/optical photons emitted by the accretion disk are absorbed by circumnuclear dust that re-emits the energy in the IR band.
Therefore, the reprocessed radiation field of the active black hole appears as thermal radiation in the (mid-)IR part of the electromagnetic spectrum. 
Heavily obscured AGN missed by X-ray observations should in principle be present in (mid-)IR surveys as it is the obscuring material itself that emits reprocessed radiation.  The main limitation in finding them is that dust clouds heated by star-formation events also emit copious amounts of energy in the IR wavelength regime. This component can dominate the thermal emission of AGN, making their identification difficult. The contrast between star-formation and AGN thermal radiation is larger in the mid-IR because of the different temperatures of the medium heated by each process. It is, therefore, easier to isolate the AGN component in the mid-IR and separate it from stellar processes in the host galaxy. This has led to studies that use the shape of the mid-IR continuum \citep[e.g.][]{Park2010},  mid-IR colour diagnostics \citep[e.g.][]{Donley2012, Messias2012,Mateos_2012, Stern, Assef2018} or template fits to multiwavelength (ultra-violet to IR) photometric observations \citep[e.g.][]{Pouliasis2020, Mountrichas2021, Thorne2022} to compile AGN samples. These can be combined with X-ray observations to search for signatures of high levels LOS obscuration that blocks the direct view to the central engine at X-rays \citep[e.g.][]{Stern2014, DelMoro2016, Hickox2018, Vito2018}. The key feature of the mid-IR observations in this type of analysis is that they are thought to provide a good proxy of the intrinsic accretion luminosity even in the case of deeply buried AGN \citep[e.g.][]{Risaliti_1999, Gandhi_2009}. Sources that appear X-ray faint for their mid-IR luminosity are obscured AGN candidates \citep{Georgantopoulos_2011}. This type of analysis suggests the presence of heavily obscured AGN in the mid-IR that are likely underrepresented in X-ray surveys. However, the selection function of mid-IR AGN samples is often complex and depends on the contrast between the  accretion luminosity and the stellar emission of the host galaxy. As a result, the calculation of the space density of the underlying population from mid-IR selected samples is not straightforward \citep[but see][]{Delvecchio2014, Assef2015}. 

This work presents a methodology that combines X-ray and mid-IR information within a Bayesian framework to constrain the space density of heavily obscured AGN. At the core of the method are X-ray observations that provide estimates of the column density $N_{\rm H}$ of individual sources and a well-understood sample selection function. IR photometry is coupled with template fits to yield estimates of the reprocessed accretion luminosity and help improve X-ray spectral constraints. Section \ref{Data_sec} describes the multiwavelength observations used in the analysis. Section \ref{Method_sec} describes the extraction and fitting of the X-ray spectra and compares the inferred X-ray spectral parameters with previous studies. Section \ref{L6improve_subsec} presents the new methodology to improve the X-ray fitting by combining it with AGN mid-IR prior information. In Section \ref{Obscured_demo_sec}, the X-ray spectral analysis results are used to infer the obscured AGN demographics and constrain the AGN space density and intrinsic CTK fraction as a function of $z$, $L_X$ and $N_{\rm H}$. We discuss our results in Section \ref{Discussion_sec}, and summarize our conclusions in Section \ref{Summary_sec}. This paper adopts a cosmology with a Hubble constant of $H_0=70$km s$^{-1}$Mpc$^{-1}$, mass density parameter, $\Omega_M=0.3$. and effective mass density of the dark matter $\Omega_\Lambda = 0.7$.

\section{Data}\label{Data_sec}

This paper uses data from the Cosmic Evolution Survey (COSMOS) field \citep{Scofield}, which benefits from a large number of multiwavelength observations over an area of nearly $\rm 2\,deg^2$. Of particular interest to this work are the \textit{Chandra} X-ray survey of this field \citep{Civano_2016} that is used to select AGN, and the multiwavelength photometry catalogue presented by  \cite{Jin_2018}. The latter is used to perform Spectral Energy Distribution (SED) templates fitting of the X-ray sources in our sample and provide independent constraints on their accretion luminosities. The various datasets used in our analysis are described below.

    \subsection{\textit{Chandra} COSMOS Legacy X-ray survey}\label{COSMOS_subsec}
    
We use X-ray data obtained as part of the \textit{Chandra} COSMOS \citep{Elvis_2009} and \textit{Chandra} COSMOS Legacy \citep{Civano_2016} survey programmes. The former consists of 36 overlapping pointings observed by the Chandra/ACIS-I detector. The latter programme provides an additional 56 Chandra/ACIS-I pointings. The two programmes together cover a total area of about 2\,deg$^2$ with a homogeneous exposure time of about 160\,ks for the inner 1.5 deg$^2$ and 80ks for the outer regions. The flux limit of the survey in the 0.5-10 keV band is $8.9 \times 10^{-16}$\,erg\,s$^{-1}$.
These observations are reduced using the analysis and source detection pipeline described by \cite{Laird_2009} and \cite{Nandra_2015}. Sources are detected independently in four spectral bands 0.5-2\,keV (soft), 2-7\,keV (hard), 4-7\,keV (ultra-hard) and 0.5-7\,keV (full). The wavelet-based source detection algorithm implemented in the {\sc wavdetect} task of the CIAO (\textit{Chandra} Interactive Analysis of Observations) data analysis system \citep{Fruscione2006} is used to provide a preliminary seed source list in each of the bands above. Photons are then extracted at the positions of these sources within apertures of variable size that correspond to the 70\%  Encircled Energy Fraction (EEF) radius of the \textit{Chandra} Point Spread Function (PSF). The background expectation value in each aperture and spectral band is also estimated after removing the contribution of source photons from the corresponding images. The extracted photons and background values at the source positions are then used to calculate the Poisson false detection probability that the observed number of photons results only from background fluctuations. A source is qualified as such if the Poisson false detection probability is $<4 \cdot 10^{-6}$ (4.5$\sigma$). There are overall 3627 X-ray sources detected in the $\sim \rm 2\,deg^2$ area of the \textit{Chandra} COSMOS Legacy survey field. Of these sources,  3372, 2772, 2140 and 971 are detected in the full, soft, hard and ultra-hard bands, respectively, as indicated in the last row of Table \ref{tab:nb_sources_per_band_and_redshift_type}. This catalogue has already been presented in \cite{Georgakakis_2017} and has been used in \cite{Aird_2017,Aird_2018, Aird_2019}. One of the motivations of our analysis is the demographics of heavily obscured AGN, and therefore the hard band sample with 2140 X-ray sources is of particular interest.

Also important for our analysis is the selection function of the X-ray sample, i.e. the probability of detecting within the \textit{Chandra} COSMOS Legacy survey an X-ray AGN with a given set of intrinsic properties (e.g. accretion luminosity, LOS obscuration, redshift). For that purpose, we use sensitivity maps generated using the methodology described in  \cite{Georgakakis_2008} and quantify the detection probability of an X-ray source with a specific photon count rate as a function of its position within the surveyed area. In later sections, we combine these sensitivity maps with AGN X-ray spectral models to link the probability of detection to the AGN intrinsic properties.

The X-ray sources are matched with their optical counterparts using different catalogues, including S-COSMOS \citep{lecolonel_2007} and the COSMOS Intermediate and Broad Band Photometry Catalogue 2008 \citep{Capak_2007}. The identification is performed using the likelihood ratio method \citep{Sutherland_1992, Brusa_2007} which takes into account the separation between the X-ray and optical position, but also the counterpart magnitude with respect to the background magnitude distribution as a prior. More details on the identification methodology applied to the COSMOS Legacy field can be found in \cite{Aird_2015}.

The compilation of spectroscopic and photometric redshifts for our X-ray sources is presented in \cite{Georgakakis_2017}. The photometric redshift estimates are primarily from \cite{Marchesi2016_IDS} based on methods presented by \cite{Salvato_2011}. For the X-ray sources in our data reduction that do not appear in the \cite{Marchesi2016_IDS} catalogue, photometric redshifts are estimated following the methods described in \cite{Aird_2015}. Among the 3627 X-ray sources, 1917 have a spectroscopic redshift, 1527 of the remaining have a photometric redshift, and 183 do not have any redshift estimation. Table \ref{tab:nb_sources_per_band_and_redshift_type} presents the number of sources detected in each X-ray band as a function of the redshift type (e.g. photometric vs spectroscopic).

\begin{table}
    \centering
    \caption{Number of sources in the  \textit{Chandra} COSMOS Legacy detected in different energy bands, full (0.5-7\,keV), soft (0.5-2\,keV), hard (2-7\,keV), ultra-hard (4-7\,keV) and any of these bands. The number of sources with spectroscopic, photometric or no redshift measurement is also shown for each subsample. The parenthesis below indicates the number of sources for each band and redshift category after applying the spatial overlap mask (see Section \ref{Jin_subsec}).}
    \begin{tabular}{c|c|c|c|c|c}
        Bands & Full & Soft & Hard & Ultra-hard & Any   \\
         & 0.5-7\,keV & 0.5-2\,keV & 2-7\,keV & 4-7\,keV & Band  \\ \hline \hline
        \textit{z} spectro  & 1825 & 1562 & 1318 & 726 & 1917  \\
                            & (1469) & (1250) & (1071) & (573)  & (1551) \\\hline
        \textit{z} photo    & 1405 & 1061 & 773  & 227 & 1527  \\
                            & (1146) & (864) & (629)  & (184) & (1259) \\\hline
        No \textit{z}       & 142  & 149  & 49   & 18  & 183  \\
                            & (119) & (122) & (44) & (17) & (155) \\\hline
        Total               & 3372 & 277)2 & 2140 & 971 & 3627  \\
                            & (2734) & (2236) & (1744) & (774) & (2965) \\
    \end{tabular}

    \label{tab:nb_sources_per_band_and_redshift_type}
\end{table}

    \subsection{COSMOS multiwavelength catalogue}\label{Jin_subsec}

We use the "super-deblended" far-IR  to (sub)millimetre photometric catalogue of the COSMOS field presented by \cite{Jin_2018}. They selected samples of galaxies detected at $\rm 2.2\mu m$, $24\,\mu \rm m$ and radio frequencies (3\,GHz) as priors for de-blending far-IR to sub-mm images from different instruments. 
They collect IR photometric information from UltraVISTA-DR2 with the kband at 2.2$\mu m$ \citep{McCracken_2012}, Spitzer IRAC at 3.6, 4.5, 5.8, and 8$\mu m$ \citep{Scoville_2007}, Spitzer MIPS at 24$\mu m$ \citep{LeFloch_2009}, Herschel PACS at 100 and 160$\mu m$ \citep{Lutz_2011}, and Herschel SPIRE at 250, 350 and 500$\mu m$ \citep{Griffin_2010}.
The IR photometry of the \cite{Jin_2018} catalogue is used to fit templates of AGN and galaxies and get a measurement of the accretion luminosity (if any) emerging in the IR part of the electromagnetic spectrum. This information is used in later sections to guide X-ray spectral fits of obscured AGN.

The spatial overlap of the \cite{Jin_2018} catalogue and the \textit{Chandra} survey of the COSMOS field is not perfect. We use the HEALPix \citep[Hierarchical Equal Area isoLatitude Pixelisation,][]{Gorski_2005} tessellation of the sky to determine the Multi-Order-Coverage \citep[MOC,][]{Fernique_2019} maps that define the irregular areas covered by two samples and assess their overlap. The maximum HEALPix order parameter for determining the MOCs is set to 14. This value corresponds to a maximum spatial resolution of about 13\,arcsec for the resulting MOCs. It is then straightforward to define the overlap region of the two MOCs and determine which X-ray sources have sky coordinates within the common area of the two samples. This reduces our sample to a total of 2965 X-ray sources, of which 2753, 2282, 1816 and 831 are detected in the full, soft, hard and ultra-hard bands, respectively.

\section{X-ray spectral analysis}\label{Method_sec}

In this section, we describe our X-ray fitting pipeline. We start by extracting the X-ray spectra while optimising for the signal-to-noise ratio (section \ref{spectral_extraction}). Then, we present the fitting algorithm (section \ref{BXA_subsec}) and the adopted X-ray AGN model (section \ref{uxclumpy_subsection}). Lastly, we justify our choice of baseline model between different setups (section \ref{difsetup_subsec}) and compare our results with previous studies (section \ref{compaLanzuisi_subsec}).

    \subsection{Spectral Extraction}\label{spectral_extraction}

The pipeline that extracts the X-ray spectra of the \textit{Chandra} detected sources consists of a collection of Python modules that call CIAO\,4.13 \citep{Fruscione2006} routines. The extraction method is similar to the one described in \cite{Marchesi_2016} but modified to maximise the Signal-to-Noise Ratio (SNR) of individual sources. 

\begin{figure}
    \centering
    \includegraphics[trim = 0cm 0cm 0cm 0cm, clip,width=0.45\textwidth]{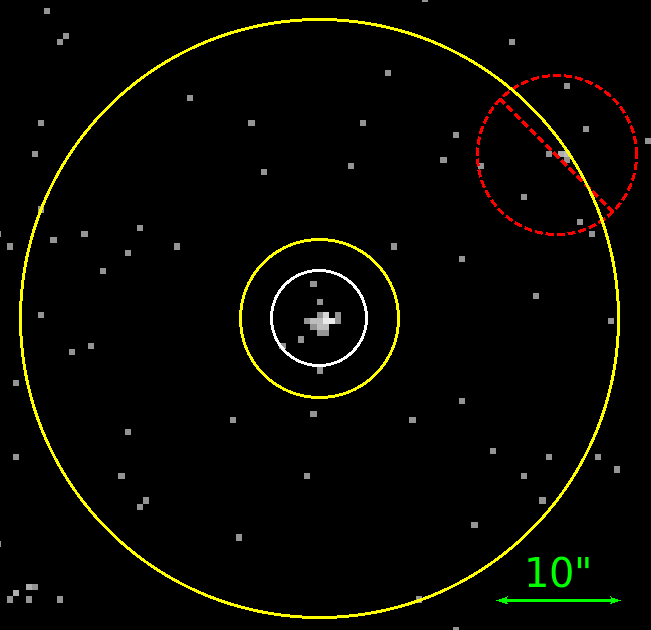}
    \caption{Example of a source and background X-ray spectral extraction regions for a radius set to EEF=70\%. The source photon extraction region is represented by the white circle with a radius of 3.80 arcsec. The background photon extraction region is defined by the yellow annulus with an inner and outer radius of 6.30 and 23.80 arcsec, respectively. A nearby X-ray source partially overlaps with the background region in this example. The region around this source, indicated by the barred red dashed circle with a 6.34 arcsec radius, is masked out before extracting the background counts.}
    \label{fig:regions}
\end{figure}

The \textit{Chandra} survey of the COSMOS field consists of many overlapping observations (see section \ref{COSMOS_subsec}). Each source is, therefore, typically present in several distinct \textit{Chandra} pointings. The number of observations per source varies between 1 to 15, with a mode of 4. The best X-ray spectrum extraction region is then determined on the stacked image of each source and is defined as
a circular region with a radius expressed in EEF that can have the value of 50, 60, 70, 80, 90, 95\%. 
The algorithm first extracts the source photons, $C_S$, within each EEF radii above. The background photons, $C_B$, are extracted from the stacked image within an annulus with an inner radius 2.5\,arcsec larger than the adopted EEF radius and with a width of 17.5\,arcsec. Any X-ray sources within the background region are masked out by excluding photons that lie within the selected EEF radius plus 2.5\,arcsec. An additional requirement is that the background region includes at least 100 counts in the stacked image. If not, the outer radius is sequentially increased by 5\,arcsecs until the condition is met. This process is graphically demonstrated in Figure\,\ref{fig:regions}.

\begin{figure}
    \centering
    \includegraphics[width=0.45\textwidth]{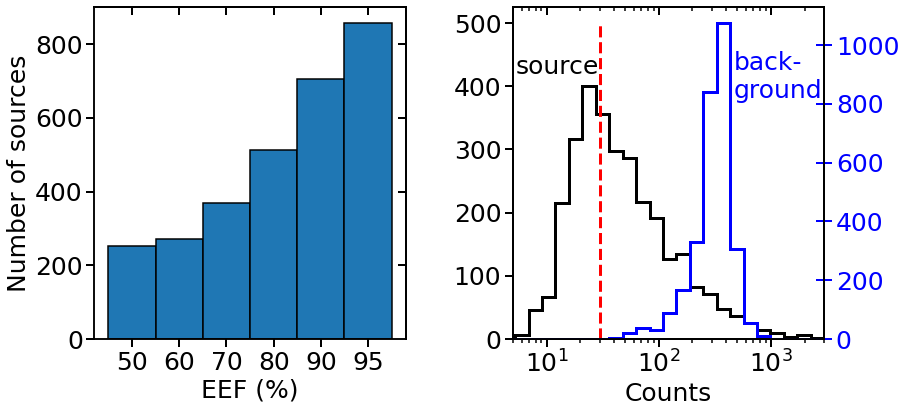}
    \caption{\textit{Left panel}: Distribution of the X-ray spectral extraction radius in EEF units for the \textit{Chandra} COSMOS Legacy sources. \textit{Right panel}: The black histogram shows the distribution of the photon counts in the full band of the extracted X-ray spectra of our sources. The red vertical dashed line represents the 30 counts limit adopted by previous studies as the threshold above which the X-ray spectra are analysed \citep{Marchesi_2016, Lanzuisi_2018}. The blue histogram shows the distribution of the photon counts in the full band of the extracted background spectra our sources.
}
    \label{fig:counts_distrib2}
\end{figure}

For each EEF value, we estimate the SNR as follows
\begin{equation}
    SNR = \frac{C_S - C_B \cdot R}{\sqrt{C_S + C_B \cdot R^2}} \, ,
\end{equation}
where R is the ratio between the areas of the source and background extraction regions. The number of counts is obtained using the \texttt{dmextract} and \texttt{get\_counts} tasks of CIAO. The EEF value that maximises the SNR in the full band is used to define the source and background extraction regions.
The distribution of the selected EEF is shown on the left panel of Figure \ref{fig:counts_distrib2}.

The extraction regions of a given source are fef into the \texttt{specextract} task to extract the source and background spectra from each Chandra/ACIS-I observation that overlaps with the position of interest. The same task also generates the corresponding Auxiliary Response Files (ARF) and Redistribution Matrix Files (RMF).
The ARF represents the efficiency of a detector as a function of the energy averaged over time. It is the product of the effective area and quantum efficiency. The effective area measures the detector's spatial size that is sensitive to photons of a given energy. The quantum efficiency measures the fraction of the incident photons that are registered by the detector as a function of the energy. The RMF describes how the energy of an incident photon is redistributed to the energy channels of the detector because of the imperfect charge collection. The ARF and RMF calibration files are necessary for the X-ray spectral analysis.
To obtain the final spectrum of a source, we combine the spectra extracted from the different observations by using the \texttt{combine\_spectra} task. The source and background ARF and RMF calibration files are also combined by weighting by exposure time.

The right panel of Figure \ref{fig:counts_distrib2} plots the distribution of the photon counts of the combined extracted X-ray spectra at the positions of \textit{Chandra} COSMOS Legacy sources. The plot also shows the distribution of the photon counts of the corresponding extracted background spectra. 
As the 100 background counts threshold is applied on the stacked image, the background count can be lower after extraction because pointings with no photon counts in the source region are not extracted, even if the background region contains photon counts.

Previous X-ray spectral studies using the \textit{Chandra} COSMOS Legacy observations have adopted a limit of 30 net counts in the X-ray spectrum for spectral analysis \citep{Marchesi_2016, Lanzuisi_2018}. In our work, no such threshold is applied. Instead, we extract and analyse the X-ray spectra of all detected \textit{Chandra} COSMOS Legacy sources. Among our 2965 X-ray sources (see Section \ref{Jin_subsec}), 1821 (61\%) have at least 30 counts, and 1141 (39\%) have less than 30 counts in the $0.5-7$keV band.

Table \ref{tab:spectral_extraction_info} is an extract of the table compiling the spectral extraction information on the 2965 sources of our sample. A full version of this table is available in electronic format. 

\begin{table*}
    \centering
    \caption{X-ray spectral extraction properties of the sources. (1) source ID; (2-3) X-ray position; (4-5) optical counterpart position; (6) EEF used for extraction in percent units; (7) source radius in arcsecond; (8) $0.5-7$keV net photon counts of the source; (9) the net photon counts of the background in all energy band; (10) flag indicating if the source is detected in the hard band (2-7keV); (11) source ID in \citetalias{Lanzuisi_2018} if cross-matched. Full table electronically available.}
    \begin{tabular}{c|c|c|c|c|c|c|c|c|c|c}
        ID   &  RA  &  Dec  &  RA\_optical &	DEC\_optical	&  EEF&	radius\_src&	cts\_057 & cts\_bkg &	hard\_flag & ID \citetalias{Lanzuisi_2018}\\ 
        (1)& (2)& (3)& (4)& (5)& (6)& (7)& (8)& (9) & (10) & (11)\\ \hline
    COSMOS\_0\_10 & 149.802 & 1.636 & 149.802 & 1.636 & 70 & 3.13 & 25 & 252 & False & -- \\
    COSMOS\_0\_100 & 149.728 & 1.719 & 149.728 & 1.719 & 90 & 6.06 & 83 & 362 & True & lid\_1186 \\
    COSMOS\_0\_102 & 149.611 & 1.746 & 149.611 & 1.746 & 80 & 3.54 & 28 & 305 & False & -- \\
    COSMOS\_0\_103 & 149.722 & 1.753 & 149.722 & 1.753 & 60 & 3.16 & 37 & 356 & False & lid\_2444 \\
    COSMOS\_0\_104 & 149.506 & 1.809 & 149.506 & 1.809 & 95 & 5.89 & 88 & 162 & True & lid\_970 \\
    \vdots & \vdots & \vdots & \vdots & \vdots & \vdots & \vdots & \vdots & \vdots & \vdots  & \vdots \\
    COSMOS\_8\_95 & 150.398 & 2.797 & 150.398 & 2.797 & 90 & 4.19 & 54 & 312 & True & lid\_427 \\
    COSMOS\_8\_96 & 150.479 & 2.798 & 150.479 & 2.798 & 95 & 5.67 & 67 & 286 & True & lid\_401 \\
    COSMOS\_8\_97 & 150.454 & 2.806 & 150.454 & 2.806 & 95 & 5.3 & 295 & 314 & True & lid\_395 \\
    COSMOS\_8\_98 & 150.515 & 2.81  & 150.514 & 2.81  & 95 & 4.42 & 64 & 195 & True & lid\_410 \\
    COSMOS\_8\_99 & 150.672 & 2.811 & 150.672 & 2.811 & 50 & 2.32 & 28 & 116 & False & lid\_487 \\

    \end{tabular}
    \label{tab:spectral_extraction_info}
\end{table*}

    \subsection{X-ray spectral fitting algorithm}\label{BXA_subsec}
The extracted X-ray spectra are fitted using the Bayesian X-ray Analysis (BXA) package presented by \cite{Buchner_2014}. We assume an observed X-ray spectrum, $D$, and a spectral model, $M$, described by a set of parameters, $\Theta$. In a Bayesian framework, the probability $\mathcal{P}(\Theta | D, M)$ of the parameter set given the observation and the model is

\begin{equation}
    \mathcal{P}(\Theta | D, M) = \frac{\Pi(\Theta | M)}{\mathcal{Z}(D | M)} \, \mathcal{L}(D | \Theta) \, ,
    \label{equa bayes}
\end{equation}
\noindent

\noindent where $\Pi(\Theta | M)$ is the prior knowledge of the parameter set for the chosen model. The model evidence, $\mathcal{Z}(D | M)$, is the probability of obtaining the observed data given the model. As it is independent of the specific parameters set of the model, it is the suited value to compare different models. This is an important feature of the Bayesian analysis that allows the selection of the model that best represents the observations. The likelihood, $\mathcal{L}(D | \Theta)$, is the probability of obtaining the observed data for the set of model parameters. Fitting an X-ray spectrum usually involves the optimization of the likelihood to yield constraints on the spectral model parameters. However, nested sampling algorithms such as MLFriends \citep{Buchner_2014b, Buchner_2019b}, explore the entirety of the parameter space at once. They first draw parameter samples from the prior distribution and then iteratively replace the lowest likelihood points with new ones drawn from the prior with a higher likelihood. The posterior distribution is constructed from the removed points weighted by their likelihood and the parameter space volume they represent. This type of algorithm is capable of exploring large parameter spaces without getting stuck in local minima, and the returned posterior distribution fully encapsulates the uncertainties of the parameter estimation. The BXA package is using the MLFriends algorithm powered by the UltraNest package \citep{Buchner_2021}. 

Ultranest works with any likelihood function. Because of the typically low number of photon counts of the X-ray spectra, in this work we estimate the likelihood using a Poisson log-likelihood function often referred to as the CSTAT statistic \citep{Cash_1979}. This is expressed as
\begin{equation}
    \mathcal{L}_{\rm CSTAT} = \, 2*\sum_i M'(i) - C(i) \left[\ln C(i) - \ln M'(i)\right]
     \, ,
    \label{equa CSTAT}
\end{equation}

\noindent where $C(i)$ is the observed number of counts in the energy bin $i$ and $M'(i)$ is the expected photon counts from the given model convolved with the RMF and multiplied by the ARF. The summation is over all the energy bins of the X-ray spectrum.

The extracted X-ray spectra at the positions of X-ray detections include contributions from both source and background photons. Therefore, the use of a Poisson likelihood for the X-ray spectral analysis requires modelling both components, i.e. including a model for both the source and background spectra. Therefore in Equations \ref{equa bayes} and \ref{equa CSTAT}, the spectral model, $M$, is the sum of the source, $M_{\rm source}$,  and background, $M_{\rm bkg}$, models. The parameter set $\Theta$ also includes the background model parameters.

    \subsection{The UXCLUMPY spectral model}\label{uxclumpy_subsection}

We choose to fit our observations with a physically-motivated X-ray spectral model that is built upon the current knowledge of the distribution of matter in the vicinity of a black hole, which describes the interaction of X-ray photons with the surrounding medium in a self-consistent way. Commonly used AGN-oriented physical X-ray spectral models include BNTORUS \citep{Brightman&nandra2011} and MYTORUS \citep{MYTORUS}, both of which assume a cylindrically-symmetric uniformly-distributed obscuring medium and describe the transmission of X-ray photons through it. In this work, we opt for the UXCLUMPY torus model presented in \cite{Buchner_2019}. It simulates a toroidal-shaped obscuring medium which is, however, clumpy by nature and hence, consistent with the observations of eclipsing events that are believed to be responsible for the observed varying obscuration in the X-ray spectra of AGN \citep[e.g.][]{Risaliti_2002}. 
The clumpiness of the obscurer is also required by mid-infrared observations to explain, among others, the diversity of the observed SEDs or the strength of the $10\mu m$ silicate feature \citep{Ramos_2009}.
The X-ray spectral model is constructed with XARS \citep{Buchner_2019}, a radiative transfer code that computes the transmission of the photons emitted by the source and interacting with its surrounding material. The photons are collected in different inclination bins representing the LOS of the observer. It returns a complex spectrum that depends on both the source's physical properties and the obscurer's geometrical parameters.

The central source emits the photons with the energy distribution of a power-law with a high energy cut-off and is surrounded by spherical obscurers distributed in toroidal geometry. These obscurers represent clouds, and to each of them is assigned a fixed density that is drawn from a log-normal distribution with mean $10^{24}$cm$^{-2}$ and a standard deviation of  1\,dex. The clouds are axisymmetrically dispersed and their number along the LOS to the observer decreases exponentially with the inclination towards the poles. Their radial distribution is uniform over 2 orders of magnitude. The radius of the clouds is distributed to reproduce the observed rate of the eclipsing events \citep{markowitz_2014}. 
Each emitted photon has a probability of escaping without interacting with the medium. Such photons correspond to the transmitted component. As the column density along the LOS increases, the non-interaction probability decreases. When interacting with the medium, the photon can either be photoelectrically absorbed or be Compton-scattered. The reflected component corresponds to all photons that have been scattered at least once. 
At a certain high energy threshold, the photoelectric absorption by neutral Fe can trigger a fluorescence process that emits photons at specific wavelengths, mainly at the FeK$\alpha$ emission line ($\sim6.4$\,keV). Photons produced this way constitute the fluorescent-line component of the UXCLUMPY model.
At energies higher than 10\,keV, an excess of photons is often observed in the curvature of the spectrum peaking at 20-30keV and is due to Compton scattering \citep{Elvis_2000}. This so-called Compton hump requires an additional obscurer that reflects the intrinsic AGN emission toward the observer without being affected by clouds of the clumpy torus \citep{Cristina_2017}.
In UXCLUMPY, this obscurer is modelled as a ring of CTK ($N_{\rm H}>10^{24}$cm$^{-2}$) clouds that are in contact with each other forming a thick doughnut-shaped structure around the central engine. 

In obscured Seyfert galaxies in the local Universe, an excess of soft X-ray photons is often observed in their spectra \citep{Bianchi_2006}. This is believed to be produced by the elastic scattering of photons emitted by the AGN onto photo-ionised gas clouds located above the obscurer and thus not interacting with it. The photons collected after this process belong to the soft or Thomson scattering component, and their spectrum mirrors the intrinsic spectrum (power-law spectral shape) emitted by the central engine.

In summary, the spectral components included in our modelling based on the UXCLUMPY implementation are (i) the transmitted X-ray component, (ii) the reflected component, (iii) the fluorescence emission lines and (iv) the soft energy excess emission. 
The three first components are merged in a single table model, whereas the Thomson scattering component is in a second and optional table. The parameters of this additional table model are linked to the parameters of the main table (i.e. redshift, power-law index of the obscurer) except for the normalisation, which is left free to vary. 
Galactic absorption is further applied to the components above. We choose to fix the Galactic column density for all sources to the average $N_{\rm H}$ value in direction of the COSMOS field, $1.72 \times 10^{20}$cm$^{-2}$ \citep{Kalberla_2005}. 

The UXCLUMPY model has several geometrical and physical parameters impacting the spectrum shape. The exponential inclination distribution of the absorbers around the central source is characterised by the TORsigma parameter varying between 0\,deg, an infinitesimally thin torus, and 84\,deg, almost a sphere. The ring of CTK absorbers is characterised by its covering factor, the CTKcover parameter, varying between 0, an infinitesimally thin equatorial disk, and 0.6, where the large equatorial CTK clouds cover 60\% of the lines of sight. The inclination angle parameter of UXCLUMPY controls the position of the observer relative to the vertical (symmetry) axis of the adopted torus geometry. The UXCLUMPY table model allows three broad bins of inclination angles 0-30\,deg (face-on),  30-60\,deg (intermediate) and 60-90\,deg (edge-on), each of which includes a wide range of LOS column densities.
Due to degeneracies with the rest of the model parameters, CTKcover and TORsigma are difficult to constrain even for high photon counts sources. In our analysis, we fix the CTKcover and TORsigma parameters to their default values, 0.4 and 28, respectively (see Table \ref{tab:parameters_torus2}).
These values are obtained by fitting the X-ray spectra of local AGN \citep{Buchner_2019}. In our baseline implementation, we further fix the inclination angle parameter to be 45\,deg, in the range intermediate inclination bin, but we also explore and quantify the sensitivity of our results to this choice.  

We fit our sample with six different setups of the X-ray spectral model. The first one is the sum of the torus model and the soft scattering component with an inclination angle of 18.2\,deg with respect to the symmetry axis (face-on). The second and third model variants are the same as above but with an inclination angle of 45 (intermediate) and 70\,deg (edge-on), respectively. We also consider three additional model setups identical to those above but without the soft scattering component. The comparison between the different model setups is discussed in detail in the following sections.

In all our model setups, the cut-off energy of the power-law emitted by the central engine is fixed to 200\,keV. During the fit, we have several free parameters to which we assign different priors (listed in Table \ref{tab:parameters_torus2}): the normalisation of the torus component with a log-uniform prior, the photon-index of the power-law with a Gaussian prior, the total LOS column density with a log-uniform prior and if necessary, the normalisation fraction of the scattering component with also a log-uniform prior. Uniform and log-uniform are neutral priors, only their upper and lower limits are significant and have been chosen to be reasonable. 
The normalisation of the soft scattering component is set to be equal, at the maximum, to $10^{-1.5}$ of the direct component normalisation. The Gaussian prior of the photon index, centred at 1.95 with a standard deviation of 0.15, is motivated by nearby AGN observations \citep{Nandra_1997}.
In the spectral fits, the redshift parameter of the UXCLUMPY model is fixed to the sources' spectroscopic redshift. If this is not available, the photometric redshift Probability Distribution Function (PDF) is provided to BXA as prior to the redshift parameter. If a source has no redshift estimation, a uniform prior between 0 and 6 is adopted.

As the spectrum is a combination of the source and background photons, the latter must also be modelled. In this work, we use the \texttt{automatic\_background()} command from BXA to model it \citep{Simmonds_2018}. This machine-learning-based approach trained itself on large X-ray surveys to derive the principal components describing the background and its variations. When called, this function fits the background spectrum with the principal component models and progressively increases their complexity. It verifies the background fitting improvement by using the Akaike information criterion. It finally returns the best background fit model that has all its parameters frozen except for the normalisation, which is added to the free-parameter set of the full model. All the parameters are indicated in Table \ref{tab:parameters_torus2}.

\begin{table}
    \centering
    \caption{Table summarizing the input parameters of the UXCLUMPY model and their prior used in our analysis. $^{a}$ the logarithmic norm of the soft scattering component corresponds in reality to log(norm\_torus) + log(norm\_scattering). $^{b}$ the init\_val correspond to the initial background normalisation value computed by the \texttt{automatic\_background()} command and re-scaled to the source area.}
    \begin{tabular}{l|l|}
        parameter & prior \\ \hline
        CTKcover & fixed = 0.4\\ 
        TORsigma & fixed =  28°\\ 
        Inclination  & fixed = 18.2° / 45° / 70°\\
        E\_cutoff & fixed = 200\,keV\\ 
        photon index & Gaussian(1.95, 0.15)\\ 
        log($\frac{N_{\rm H}}{cm^{-2}}$)   & uniform(20, 26)\\ 
        log(norm\_torus) & uniform(-8, 3) \\
        log(norm\_scattering)$^{a}$  & uniform(-7, -1.5)\\
        redshift spectroscopic & fixed \\ 
        redshift photometric & photometric PDF \\
        no redshift estimation & uniform(0, 6) \\
        log(norm\_background)  & uniform(init\_val$^{b}$ $\pm$2 )
        
    \end{tabular}

    \label{tab:parameters_torus2}
\end{table}

\begin{figure}
    \centering
    \includegraphics[width=0.45\textwidth]{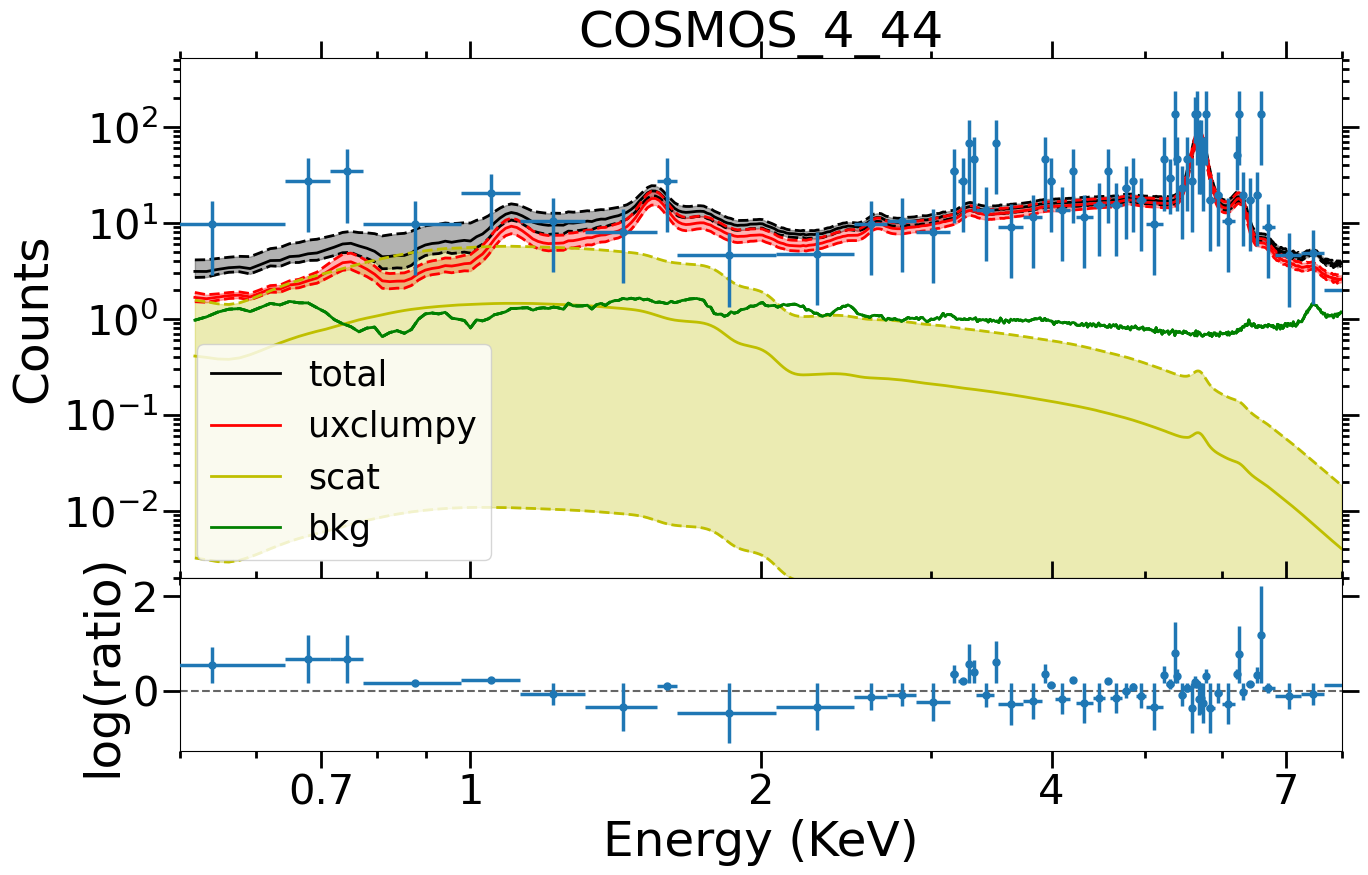}
    \caption{Example of X-ray spectral analysis result using BXA with the UXCLUMPY model. 
    The blue crosses are the extracted X-ray spectrum grouped to yield an SNR above 1 per bin. 
    The red line corresponds to the UXCLUMPY model, the yellow line represents the soft scattering, the green line is the background model. The sum of all three components above is shown with the black line. The shaded regions correspond to the 1$\sigma$ confidence interval of the corresponding component. The lower panel plots the logarithmic ratio between the X-ray spectrum and the best-fit model as a function of the energy.}
    \label{fig:fit_example}
\end{figure}

By fitting the extracted spectra between 0.5 and 8keV using BXA with UXCLUMPY as the model, we constrain the free parameters in a Bayesian framework. The choice of energy limits does not impact significantly the results. Figure \ref{fig:fit_example} shows an X-ray source of our catalogue with its best-fit model, and its different components. The error margins are computed from the parameter posterior distribution and represent their 1$\sigma$ variation.
The choice of the source is primarily justified by its relatively high photon counts, 116, in the 0.5-7keV band, and by its significant scattering component.

    \subsection{Comparison between different setups}\label{difsetup_subsec}
    
We explore the six setups of the UXCLUMPY X-ray spectral model with different fixed parameters and components described in section \ref{uxclumpy_subsection} (see Table \ref{tab:parameters_torus2}), to assess their impact on the final results and  determine the one to be used as the baseline model further in our analysis.

As discussed in section \ref{BXA_subsec} the model evidence $\mathcal{Z}$ is the best tool to compare two models with the same data set. The larger it is the more favoured is the corresponding model.
We adopt a logarithmic evidence difference $\Delta \log\mathcal{Z} > 4.6$ as the threshold to select strongly favoured models \citep{Jeffreys1961}. Additionally, we consider that a difference below this threshold still favours the highest evidence model but we cannot rule out the alternative model.

\begin{table}
    \centering
    \caption{Table presenting the evidence comparison between the different model setups relative to our baseline model that includes Thomson scattering and has an inclination angle of i=45°.
    The second column is the ensemble evidence difference relative to the baseline model. The ensemble evidence of a given model is defined as the sum of the evidences of individual X-ray sources and is listed in brackets. The error bars are estimated by bootstrapping (see text for details). The third column displays the number of individual sources having a higher evidence than the baseline model and in parentheses is the number of sources for which the model is strongly favoured over the baseline model. Similarly, the last column indicates the number of individual sources with lower evidence, and in parentheses, the number of sources for which the model can be ruled out.}

    \begin{tabular}{c|c|c|c}
        Models & Sum difference & $\Delta\log\mathcal{Z} \ge 0$ & $\Delta\log\mathcal{Z} < 0$ \\
        & (Sum total) & ($\Delta\log\mathcal{Z} > 4.6 $) & ($\Delta\log\mathcal{Z} < -4.6$) \\ \hline
        
        i=18.2°          & 18.7 $\pm$ 12.3 & 1514 & 1451 \\
        scattering     & (-1120795.3)    & (0)  & (0)\\ \hline
        i=18.2°          & -153.7$\pm$41.7 & 1426 & 1539 \\
        no scattering  & (-1120965.7)    & (0)  & (4) \\ \hline
        
        i=45°          & 0               & 2965 &  0\\
        scattering     & (-1120814.3)    & (0)  & (0)\\ \hline
        i=45°          & -201.5$\pm$39.7 & 1353 & 1612 \\
        no scattering  & (-1121013.0)    & (0)  & (5)\\ \hline

        i=70°          & -40.7$\pm$11.3  & 1389 & 1576 \\
        scattering     & (-1120855.8)    & (0)  & (0)\\ \hline
        i=70°          & -256.3$\pm$42.6 & 1303 & 1662 \\
        no scattering  & (-1121080.5)    & (0)  & (7)\\ 

    \end{tabular}

    \label{tab:favored_models}
\end{table}

Table \ref{tab:favored_models} presents the comparison of the different UXCLUMPY X-ray spectral model setups. For individual sources we estimate the logarithmic evidence difference, $\Delta\log\mathcal{Z}$, between each model setup and the one with soft-scattering and an inclination angle of 45\,deg (baseline model). For a given model setup we sum up the logarithmic evidence differences of all the sources \citep[following][]{Buchner_2014} and compute its uncertainty using a bootstrap resampling method.
It consists of taking N sources with replacement with N being the number of sources in our sample and calculating the sum of the logarithmic evidence differences relative to the baseline model. This process is repeated 100 times for each model setup. The standard deviation of $\Delta\log\mathcal{Z}$ estimated from these 100 trials represents the uncertainty of the logarithmic evidence difference of a given model setup. The table also shows the total number of individual X-ray spectra with $\Delta \log\mathcal{Z} > 4.6$ or $\Delta \log\mathcal{Z} < -4.6$ relative to the baseline model, in other words, how many times a model setup is strongly favoured or disfavoured relative to the baseline model. 

Table \ref{tab:favored_models} shows that the Thomson scattering component is strongly favoured by the data. The total logarithmic evidence of the model setups with the scattering component are larger by typically 100 compared to the same model without the scattering. Moreover, the 18.2\,deg inclination is slightly favoured with a $\log\mathcal{Z}$ difference of $18.7\pm12.3$, although the uncertainty is large.  The 70\,deg inclination is performing worse with a $\log\mathcal{Z}$ difference of $-40.7\pm11.3$. 

\begin{figure}
    \centering
    \includegraphics[width=0.45\textwidth]{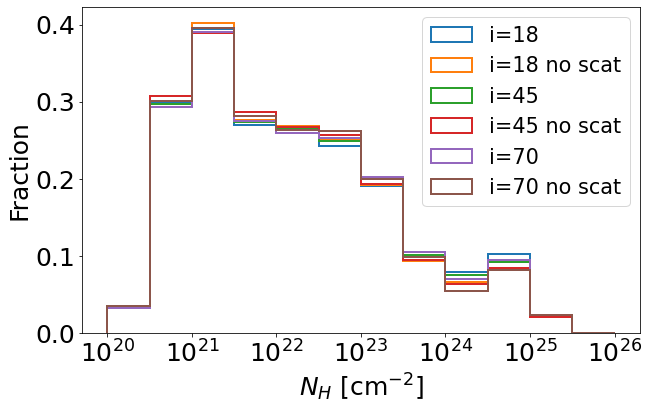}
    \caption{$N_{\rm H}$ distribution for different inclination angles and scattering parameters in the UXCLUMPY model. There are no significant differences between the different model setups.}.
    \label{fig:distribution_NH_models}
\end{figure}

\begin{table}
    \centering
    \caption{CTK sources for each X-ray spectral model setup. The first and second column provides a description of the model setup. 
    The third column is the number of sources with median posterior $N_{\rm H}> \rm 10^{24}\, cm^{-2}$ and, in parenthesis, the fraction of the total sample it represents.
    Finally, the last column shows the averaged fraction of the $N_{\rm H}$ posterior distributions above the CTK limit.}
    \begin{tabular}{c|c|c|c}
        inclination &  soft & number of   & fraction of \\
        (deg)   & scattering & CTK sources   & CTK chains (\%) \\ \hline
        
        18.2&  yes   & 302 (10.2\%) & 10.9 \\\hline
        18.2& no  & 266 (9.0\%)  & 9.2 \\ \hline
        45&  yes   & 280 (9.4\%)  & 10.5 \\ \hline
        45& no  & 251 (8.5\%)  & 8.6 \\\hline
        70&  yes   & 281 (9.5\%)  & 10.5 \\ \hline
        70& no  & 238 (8.0\%)  & 8.2 \\
    \end{tabular}
    \label{tab:CT_fraction}
\end{table}

We compare the column density distributions obtained by the different model setups in Figure \ref{fig:distribution_NH_models}. These histograms are constructed using the median of the corresponding $N_{\rm H}$ posterior distribution for each source. The resulting $N_{\rm H}$ distribution is not sensitive to the adopted model setup. Table \ref{tab:CT_fraction} further explores differences in the total number of CTK AGN among the diverse model setups. The number of CTK sources decreases slightly with the increasing inclination angle, but overall, all six models yield similar numbers of CTK sources. Moreover, most of these CTK sources (a total of 222) are common in all six model setups. 
The fraction of CTK sources also decreases by ~1\% when the model does not have a scattering component. 
One can also look at the posterior distribution instead of the median values to get a more nuanced estimation of the Compton-thickness of the sample. 
We simply average the fraction of the posterior distribution in the CTK regime for each source.
We see that the similarities among the different model setups are maintained. 

Based on the analysis above, our baseline X-ray spectral model includes a soft-scattering component, which is strongly favoured by the data, and assumes an inclination angle fixed to 45\,deg.  Although the evidence analysis shows a weak preference for lower inclinations angles (18.2\,deg; nearly face-on orientation), we opt for the intermediate group of sight-lines in UXCLUMPY (the 30-60\,deg bin) that probe a wider range of columns densities between the central engine and the observer. Figure \ref{fig:distribution_NH_models} nevertheless demonstrates that our results are not sensitive to the inclination angle choice.

    \subsection{Comparison with previous studies}\label{compaLanzuisi_subsec}

This section compares the physical parameters inferred by our X-ray spectral analysis with those derived in previous studies using the \textit{Chandra} Legacy data. Since one of the main motivations of our work is the characterization of the LOS obscuration of AGN, we limit this comparison to the hydrogen column density. \cite{Marchesi_2016} fit the X-ray spectra of \textit{Chandra} Legacy sources with more than 30 net counts in the 0.5-7 keV band using a power-law spectral model modified by photoelectric absorption. The adopted model is valid for moderately obscured AGN ($N_{\rm H}\la \rm10^{23}\,cm^{-2}$) but becomes increasingly inaccurate for higher levels of obscuration. This is because of the increasing importance of the Compton scattering for $N_{\rm H}\ga \rm10^{23}\,cm^{-2}$ and degeneracies between the fitted parameters, i.e. power-law photon index and absorbing column density. \citet[][\citetalias{Lanzuisi_2018} hereafter]{Lanzuisi_2018}  updates the spectral analysis results of \cite{Marchesi_2016} for sources that show evidence for high levels of LOS obscuration. AGN with an estimated spectral index $<1.4$ or a hydrogen column density $N_{\rm H}>\rm 10^{23}cm^{-2}$ are selected. This subsample is refitted using the MYTORUS physically-motivated model \citep{MYTORUS} that includes processes such as Compton scattering and fluorescent line emission assuming a toroidal obscurer geometry. 
\citetalias{Lanzuisi_2018} presents the physical properties of 1832 sources within the \textit{Chandra} Legacy survey. It includes the original results of \cite{Marchesi_2016} updated with the MYTORUS model fitting results for the obscured candidates.
This catalogue is cross-matched with ours within 1.7\,arcsec to yield a total of 1805 common sources. A larger radius would not significantly increase the source numbers (1814 sources at 2 arcsec) and would lead to source misidentification with several sources matched to the same source.

\begin{figure}
    \centering
    \includegraphics[width=0.45\textwidth]{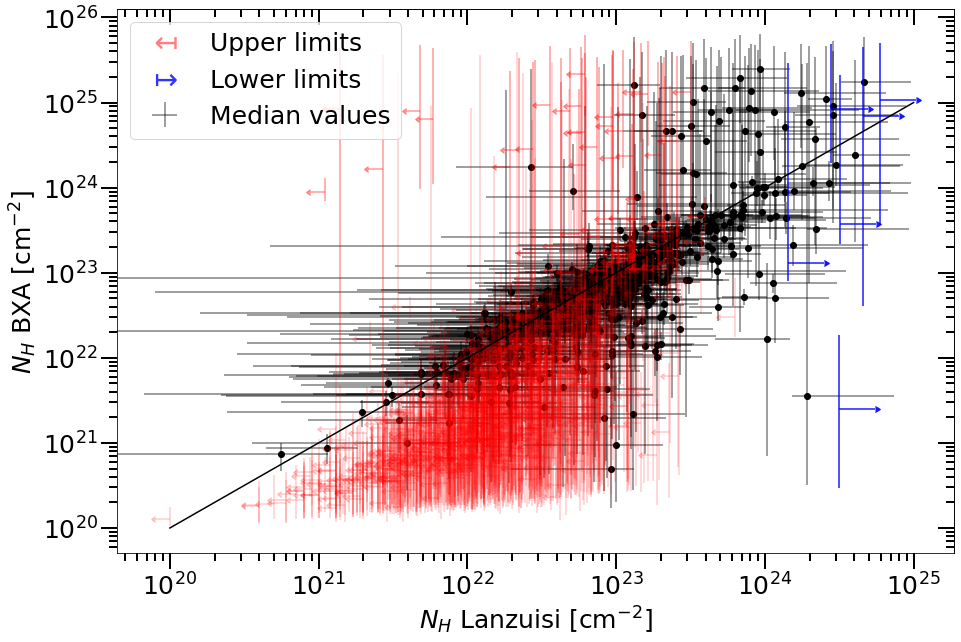}
    \caption{$N_{\rm H}$ values from our model compared to the best-fit $N_{\rm H}$ values from \citetalias{Lanzuisi_2018}. The black dots represent the median values of our BXA fit estimations and their errorbars represent their $1\sigma$ uncertainties. The red (blue) arrows are upper (lower) $N_{\rm H}$ estimates in \citetalias{Lanzuisi_2018}.}
    \label{fig:compa_Marchesi}
\end{figure}

Figure \ref{fig:compa_Marchesi} compares the $N_{\rm H}$ values obtained for our baseline model using BXA with those from \citetalias{Lanzuisi_2018}.
At moderate obscuration, there is overall good agreement between the independently estimated $N_{\rm H}$ values. The \citetalias{Lanzuisi_2018} spectral catalogue also includes a large number of AGN for which the absorbed power-law fits of \cite{Marchesi_2016} yield an upper limit to the column density. The bulk of these sources are associated with unobscured AGN in our spectral analysis, with column densities $N_{\rm H} \rm \la  10^{22} \, cm^{-2}$. 
Figure \ref{fig:compa_Marchesi} further shows that it is in the CTK regime that the most significant discrepancies between our results and those of \citetalias{Lanzuisi_2018} appear. 

\begin{figure}
    \centering
    \includegraphics[width=0.45\textwidth]{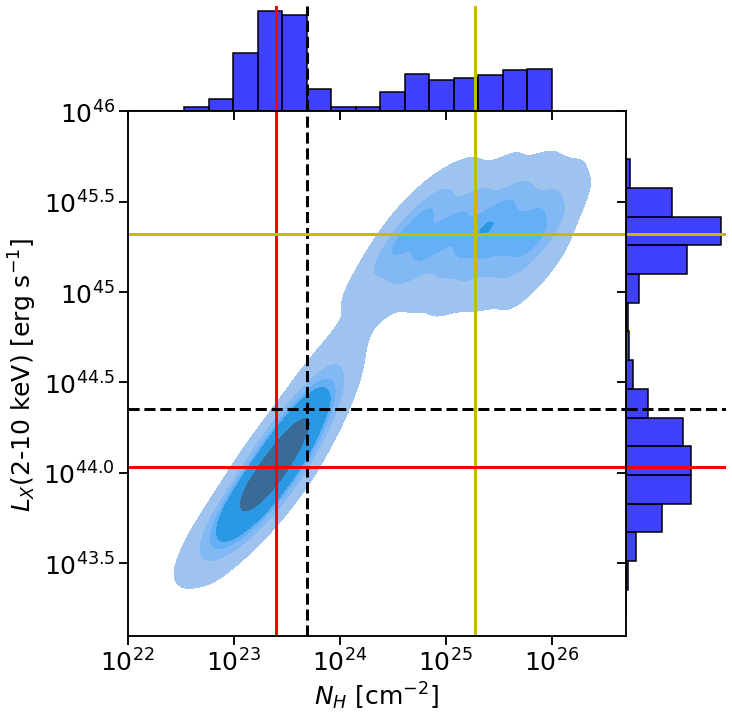}
    \caption{Density plot of the posterior probability distribution of the X-ray luminosity $\log L_X(\rm 2-10\,keV)$ as a function of the column density $\log N_{\rm H}$ for the source COSMOS\_1\_420. The 1-dimensional projections of the posterior on the X-ray luminosity and column density axes are shown on the right and top panels, respectively. The median values are plotted in dashed black lines, whereas the CTN and CTK median solutions are plotted with the red and yellow solid lines, respectively.
    }
    \label{fig:double_peaks_Lx_NH_scatter}
\end{figure}


There are 34 \citetalias{Lanzuisi_2018} sources with best-fitting $N_{\rm H}$ in the CTK regime compared to 75 AGN in our analysis with median hydrogen column densities distribution $N_{\rm H} > \rm 10^{24} \, cm^{-2}$(19 in common).  Many of the 75 sources have broad $N_{\rm H}$ PDFs that extend below $10^{24}$cm$^{-2}$  into the CTN regime but still heavily obscured.  As an example, the $N_{\rm H}$ PDF of the source COSMOS\_1\_420 is represented by the upper histogram in Figure \ref{fig:double_peaks_Lx_NH_scatter}. It shows two distinct peaks, one below $N_{\rm H}=\rm 10^{24}\,cm^{-2}$ and a flatter one above this limit. Multinested-sampling algorithms like Ultranest used by BXA allow us to explore such posterior distributions efficiently. Standard Markov Chain Monte Carlo algorithms may get stuck in one of the local minima and hence yield unimodal posteriors that do not represent the complexity of the system.

\begin{figure}
    \centering
    \includegraphics[width=0.45\textwidth]{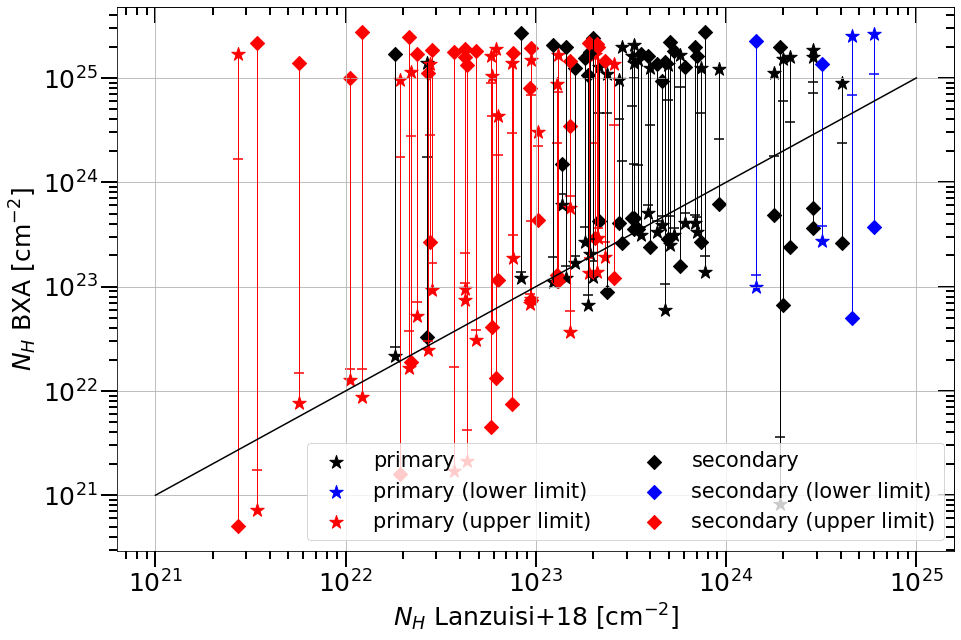}
    \caption{Comparison between the best-fit $N_{\rm H}$ determined by \citetalias{Lanzuisi_2018} and our estimates for AGN with a bi-modal $N_{\rm H}$ posterior PDF in our analysis. For each source, both peaks are plotted and connected with a line. The symbols associated with each peak represent the likelihood of that solution, i.e. the fraction of the posterior associated with the peak in question. Stars represent primary peaks (more likely), whereas diamonds indicate the secondary (less likely) peaks of the posterior. The horizontal bars on the lines connecting two peaks indicate the median from the $N_{\rm H}$ posteriors. The red and blue colours represent sources that have upper and lower  $N_{\rm H}$  limits in \citetalias{Lanzuisi_2018}, respectively. The black colours correspond to the best fit $N_{\rm H}$ values in \citetalias{Lanzuisi_2018}.}
    \label{fig:double_peaks}
\end{figure}

Point parameter estimates, like the median $N_{\rm H}$ value plotted in Figure \ref{fig:compa_Marchesi} are problematic in the case of a multi-modal PDF (e.g. Figure \ref{fig:double_peaks_Lx_NH_scatter}). Instead, for this class of sources it is more instructive to show both peaks of the PDF. We first define double-peaked sources as those for which the posterior $N_{\rm H}$ distribution includes at least 25\% CTK and CTN solutions. For these sources, both $N_{\rm H}$ peaks (connected with a line) are compared in Figure \ref{fig:double_peaks} with the best-fit $N_{\rm H}$ inferred by \citetalias{Lanzuisi_2018}. In most cases, the one-to-one relation is bracketed by two peaks of the posterior distribution function. 
We also notice that for a fraction of the double-peaked sources, the CTK part of the posterior distribution function results from small count statistics as 66\% of the double-peaked sources have less than 30 counts in the full band. The bi-modality of the posterior distribution can also be an effect of spectral model degeneracies, like the level of obscuration and the intrinsic X-ray luminosity.
One option to break the degeneracies would be to use multiwavelength information to add parameter priors into the spectral analysis to better constrain the physical properties of the sources. This approach is presented in the next section.

\section{X-ray spectral fitting improvement with multiwavelength information}\label{L6improve_subsec}

In the previous section, several sources were identified with broad or multi-modal column density posterior distributions, resulting from degeneracies between model parameters and small photon statistics. Figure \ref{fig:double_peaks_Lx_NH_scatter} plots the posterior distribution in the 2-dimensional space of X-ray luminosity and obscuring hydrogen column density for the double-peaked source COSMOS\_1\_420. The bi-modality in the column density posterior distribution is also seen in the X-ray luminosity PDF of this source. There is a strong positive correlation between these two parameters in the sense that the inferred $L_X(\rm 2-10\,keV)$ increases with increasing column density. Indeed, a higher column density requires a higher intrinsic accretion luminosity to compensate for the stronger photon absorption and reproduce the observed source flux. An independent estimate of the intrinsic AGN luminosity could therefore provide additional constraints on the X-ray spectral analysis posterior distributions and help break the degeneracies shown in Figure \ref{fig:double_peaks_Lx_NH_scatter}. 

The intuition that X-ray and IR luminosities and LOS obscuration are correlated is not recent, numerous studies in the last 30 years attempted to identify heavily obscured AGN  by looking for sources that appear X-ray underluminous for the mid-infrared or optical emission \citep{Risaliti_1999, Alexander_2008, Georgantopoulos_2011}. We build upon previous studies but instead of applying  strict and arbitrary cuts on the X-ray and IR luminosities ratio to dictate whether a source is CTK or not, we are using the IR luminosity only as a prior in our X-ray spectral analysis. For that purpose, we use the mid-IR $\nu F_{\nu}$ luminosity at $6\mu m$ to estimate the intrinsic (unabsorbed) accretion luminosity that is then converted to the intrinsic X-ray luminosity via well-established correlations \citep[e.g.][]{Stern, Mateos, Chen_2017}. The IR-derived X-ray luminosity and uncertainties are then used as a prior to guide the X-ray spectroscopy and improve constraints on the measured obscuration. Our Bayesian approach combines the uncertainties of each measure and proxy relationships in a consistent way throughout our analysis, improving the confidence of our results.

The next sections describe how templates fits to the observed SEDs of \textit{Chandra} COSMOS Legacy sources are used to constrain the AGN mid-IR luminosity and how this information feeds back to the X-ray spectral analysis to break parameter degeneracies.

    \subsection{SED fitting methodology}\label{SEDfit_subsec}

SED template fitting is an efficient tool to recover the emission associated with the AGN accretion luminosity, even when the AGN component does not dominate the SED. Observations at mid- and far-IR wavelengths are necessary to separate the AGN emission from the thermal radiation produced by stellar processes.
Therefore, we cross-match the positions of the optical counterparts of the \textit{Chandra} COSMOS Legacy survey X-ray sources with the "super-deblended" far-IR to (sub)millimetre photometric catalogue of \citet[][see section \ref{Jin_subsec}]{Jin_2018}. In this exercise, a matching radius of 1.3\,arcsec is adopted. For the density of IR sources ($\sim 10^{5}\,\rm deg^{-2}$), this search radius corresponds to a spurious fraction of 5.1\%. Among the 2965 X-ray sources in our sample, 164 have no counterpart in the \cite{Jin_2018} catalogue. From the remaining 2801 IR associations, 9 sources have redshifts above 4, which is the upper redshift limit of the SED fitting algorithm. These sources have not been analysed and are excluded from the sample. This leaves a total of 2792 X-ray sources of \textit{Chandra} COSMOS Legacy that have been matched  with \citet{Jin_2018} counterparts and fulfil the redshift criterion for the SED fit.

\begin{figure}
    \centering
    \includegraphics[trim = 0.3cm 0.4cm 0.5cm 1.4cm, clip,width=0.45\textwidth]{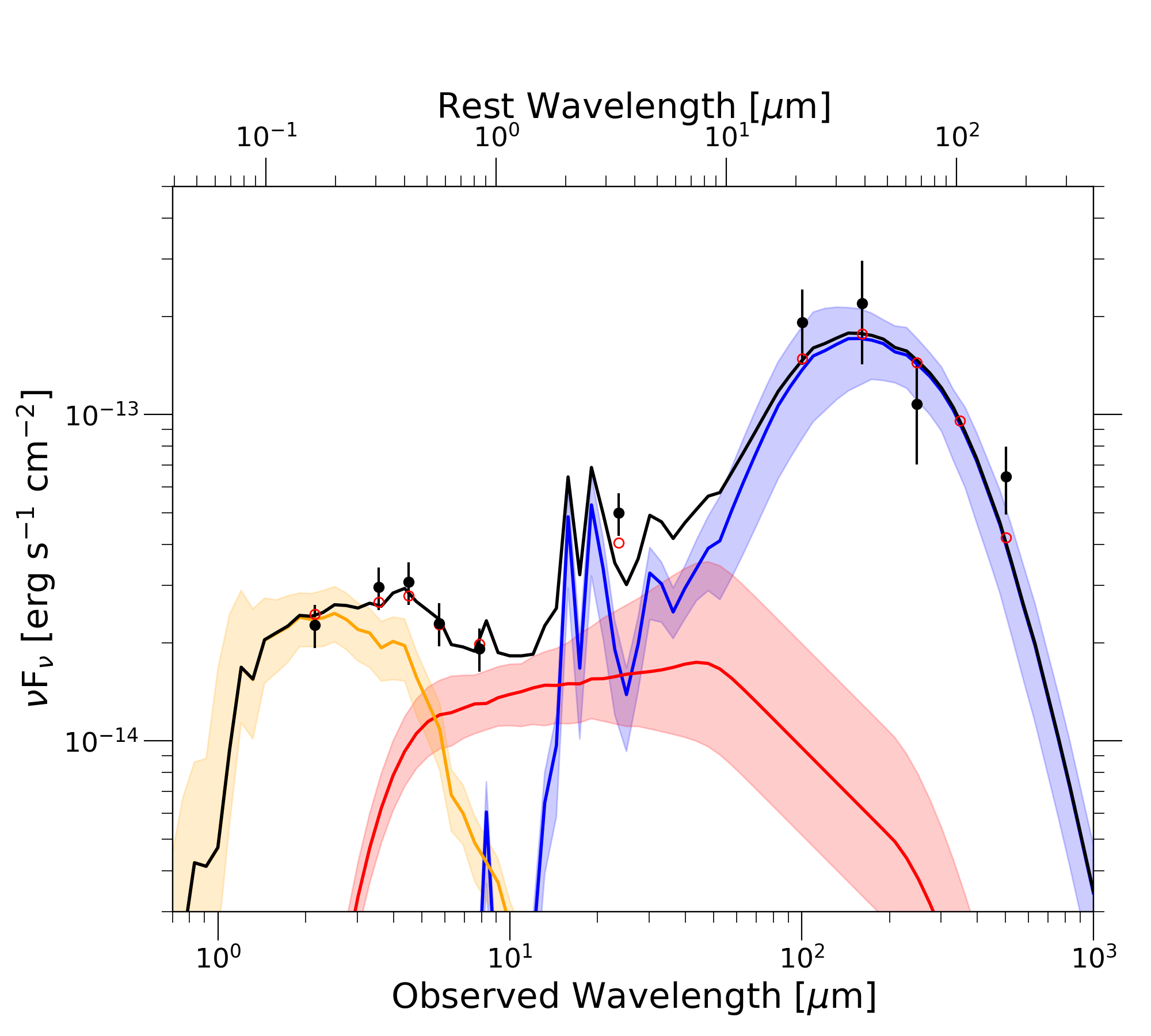}
    \caption{Example of a template fit to the IR SED of the X-ray source COSMOS\_1\_420 (object with ID 10178753 in the \protect\cite{Jin_2018} catalogue) using the FortesFit code. The black dots and vertical lines represent the photometry of the source and its uncertainties. The curves represent the stellar population emission (orange), the IR AGN emission (red), and the IR emission from star formation (blue). The shaded regions represent the approximate $1\sigma$-scatter of each SED component as constrained by FortesFit. The black curve corresponds to the sum of the above three components.}
    \label{fig:IR_SED}
\end{figure}

We fit the $2-500 \: \mu \rm m$ SED of the matching sources using the multi-component Bayesian SED fitting package FortesFit \footnote{https://github.com/vikalibrate/FortesFit} \citep{Rosario_2019}.
Our choice of fitting algorithm is motivated by its Bayesian nature. Indeed, the decomposition of the AGN and star-formation contributions in the IR part of the SED is not trivial depending  on the choice of templates to use and degeneracies between model parameters. However, FortesFit and its Bayesian inference methodology are designed to tackle these issues and capture the uncertainties of the inferred parameters. These uncertainties are then consistently propagated to the X-ray spectral analysis. Moreover, we measure the IR luminosity at 6$\mu m$, where the contrast between the host galaxy star-formation and the AGN torus IR emissions is typically large \citep[e.g.][]{Nardini_2008, Nardini_2009}, thereby facilitating the decomposition. We also choose to use state-of-the-art observationally-motivated model templates that are able to capture the observed diversity of AGN and star-formation emission in the infrared.
Our SED modelling includes the unabsorbed stellar emission from \cite{2003BC}, the IR emission coming from the torus based on the empirical DECOMPIR AGN model \citep{2011Mullaney}, and the IR emission from the dust-obscured star-formation \citep{2014Dale}. The stellar emission template has the age and mass of the stellar population as free parameters. 
As our aim is not to constrain the stellar population emission but only to consider its possible excess at rest-frame IR wavelengths that could affect the fit of the AGN emission, we do not include optical/UV photometry in our SED fits. Including optical/UV photometry does not significantly impact the AGN emission constraints and the final results within the error margins.
The free parameters of the star-forming galaxy model are the $8-1000 \: \mu \rm m$ galaxy luminosity from star formation, $L_{\rm SF}$, and a shape parameter that describes a wide range of spectral shapes for normal star-forming galaxies. The DECOMPIR template combines a broken power-law and a black body. The free parameters are the IR AGN luminosity in the interval $8-1000 \: \mu \rm m$ and the short-wavelength slope, which we vary freely in the [-0.3, 0.8] range following \cite{2011Mullaney}.
For more details on the SED fitting process, we refer to \cite{Andonie_2022b}.
Figure\,\ref{fig:IR_SED} shows the SED template fit of the double-peaked X-ray AGN previously shown in Figure\, \ref{fig:double_peaks_Lx_NH_scatter}. The emission contributions of the stellar population, AGN torus and star formation and their $1\sigma$\,uncertainties estimated by FortesFit are over-plotted in the figure.

    \subsection{Integration of the \texorpdfstring{$L_{6\mu m}$}{TEXT}  measurements into the X-ray spectral analysis}

The FortesFit SED code samples the $L_{6\mu m}$ luminosity PDF at the 1st, 16th, 50th, 84th and 99th percentiles. These point estimates are linearly interpolated to reconstruct the $L_{6\mu m}$ luminosity PDFs of the individual AGN of the sample. 
We caution that the SED fitting approach becomes less efficient in constraining the intrinsic AGN properties if they are weak relative to the stellar emission of the host galaxy. 
During the SED fit, the AGN $L_{8-1000\mu m}$ luminosity is a free parameter that has a lower limit of $\rm 10^{38} erg\,s^{-1}$.
Many X-ray sources have posterior $L_{8-1000\mu m}$ AGN luminosities that are skewed to this lower boundary. For these sources, the AGN component is essentially not needed to fit the observed SED or, equivalently, the AGN template has a much lower normalisation than the dusty star-formation component. In these cases, we use the posterior distribution to determine the $3\sigma$ upper limit to the intrinsic AGN luminosity. The empirical criterion adopted to identify such sources is that the 1st percentile of the posterior is lower than $2\times \rm 10^{38} erg\,s^{-1}$. With this criterion, among the 2792 sources with an SED fit, 1367 are constrained, and 1425 are assigned $3\sigma$ upper limits.

\begin{figure}
    \centering
    \includegraphics[width=0.45\textwidth]{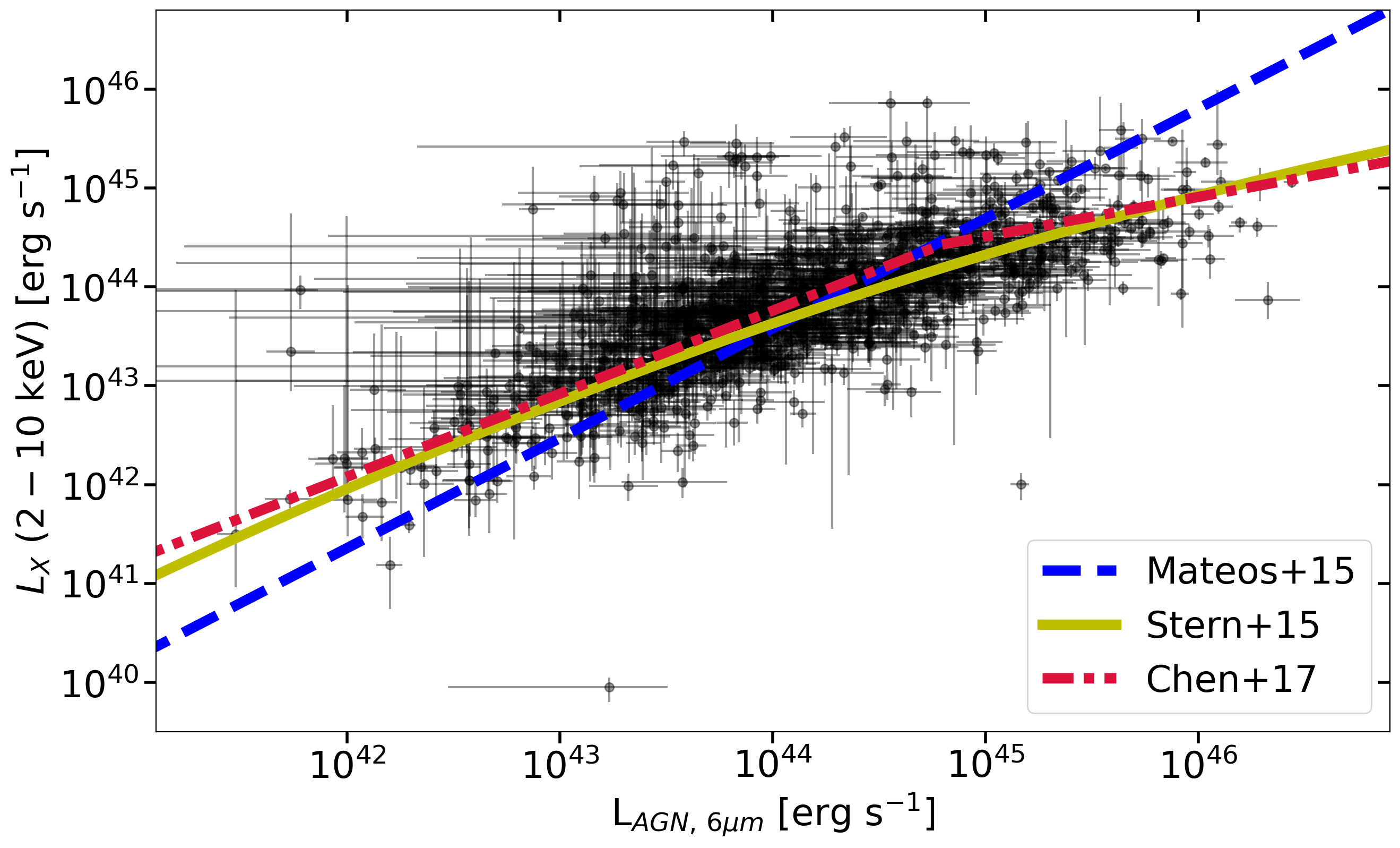}
    \caption{AGN X-ray luminosity in the 2--10\,keV band (obscuration corrected) vs the AGN $\nu F_{\nu}$ luminosity at $6\mu m$. The latter is estimated from the template fits to the observed SED of the X-ray sources in the  {\it Chandra} COSMOS Legacy field. We choose not to plot the $6\mu m$ luminosity upper limits for the sake of clarity. Sources with multi-modal X-ray luminosity posteriors are also not plotted for the same reason. 
    The dashed blue, solid yellow and dash-dotted red lines correspond respectively to the \protect\cite{Mateos}, \protect\cite{Stern}, and \protect\cite{Chen_2017} $L_X(\rm 2-10\,keV)$-$L_{6\mu m}$ relationships.}
    \label{fig:L6_Lx_plot}
\end{figure}

Figure \ref{fig:L6_Lx_plot} shows the $L_X(\rm 2-10\,keV)$ as a function of the $L_{6\mu m}$ for the X-ray sources in the {\it Chandra} COSMOS Legacy survey. As already demonstrated in previous studies, these two luminosities are well correlated indicating that $L_{6\mu m}$ can be used as a proxy for the intrinsic $L_X(\rm 2-10\,keV)$. Rather than deriving a relation between the X-ray luminosity and the $6\mu m$ luminosity from the data plotted in Figure \ref{fig:L6_Lx_plot}, we choose to use published relations and test which one describes best our observations. Three recent parametrisations for the $L_{6\mu m}-L_X(\rm 2-10\,keV)$ correlation are shown in Figure \ref{fig:L6_Lx_plot}. These curves correspond to Equation \ref{eq:Lx_L6_Chen} below for the \cite{Chen_2017} relation, Equation \ref{eq:Lx_L6_Mateos} in the case of the \cite{Mateos} work and Equation \ref{eq:Lx_L6_Stern} for the  \cite{Stern} sample:

\begin{equation}
    l_x(l_{6\mu m}) = \left\{
    \begin{array}{ll}
        0.84 \cdot (l_{6\mu m} -45) + 44.6 & \mbox{if } l_{6\mu m} \leq 44.79 \\
        0.40 \cdot (l_{6\mu m} -45) + 44.51 & \mbox{if } l_{6\mu m} > 44.79,
    \end{array}
\right.
\label{eq:Lx_L6_Chen}
\end{equation}
\begin{equation}
    l_x(l_{6\mu m}) = 0.377 + 0.90 \cdot (l_{6\mu m} -44) +44\,,
    \label{eq:Lx_L6_Mateos}
\end{equation}
\begin{equation}
    l_x(l_{6\mu m}) = 40.981 + 1.024 \cdot (l_{6\mu m} -41) - 0.047 \cdot (l_{6\mu m} -41)^2\, ,
    \label{eq:Lx_L6_Stern}
\end{equation}

\noindent with $l_X = \rm{ log}(L_X \,\rm erg\, s^{-1})$ and $l_{6\mu m} = \rm{log}(L_{6\mu m} \,\rm erg\, s^{-1})$.

\begin{figure}
    \centering
    \includegraphics[width=0.45\textwidth]{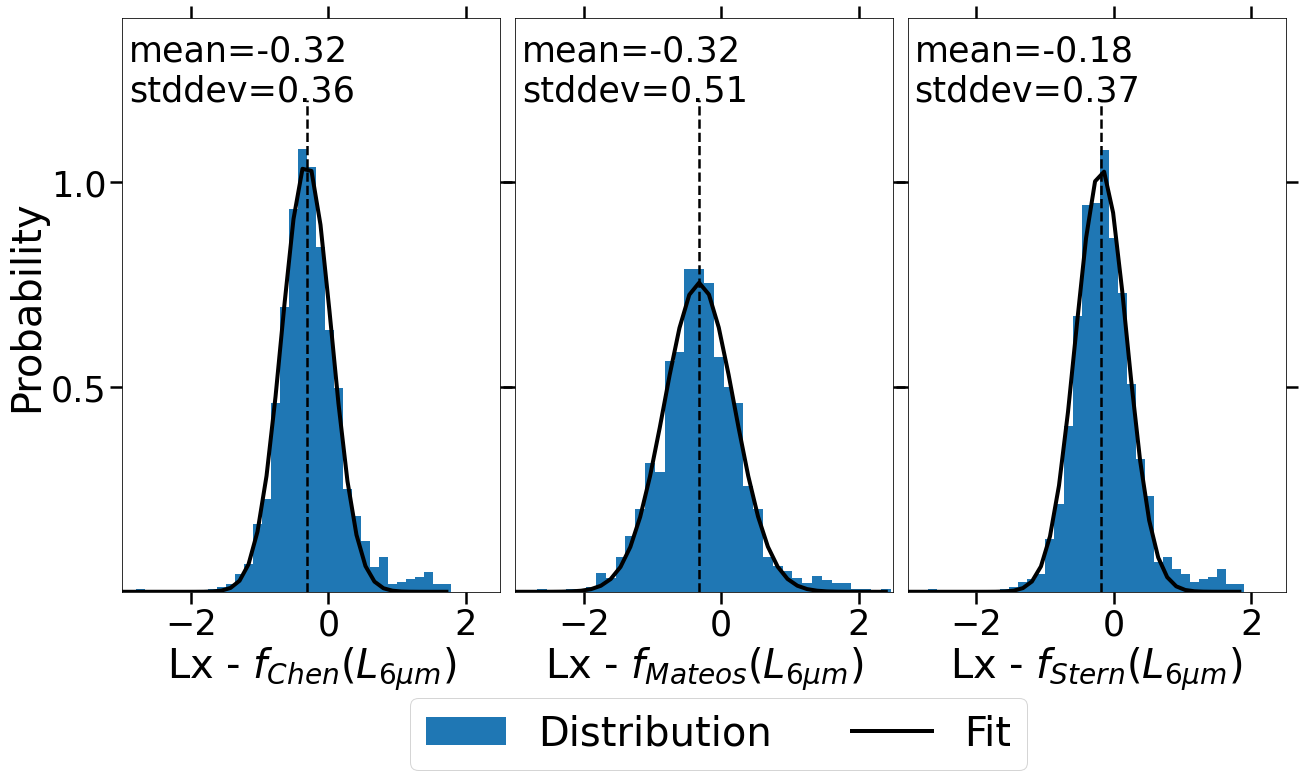}
    \caption{Distribution of the difference between the median log$L_X(\rm 2-10\,keV)$ of the posterior distribution derived by the X-ray spectral analysis and the expected log$L_X(\rm 2-10\,keV)$ determined from the AGN $L_{6\mu m}$ using the relationships of  \protect\cite{Chen_2017} (left panel), \protect\cite{Mateos} (middle panel) and \protect\cite{Stern} (right panel). The best-fit normal distributions are also shown (black curves), with their mean represented by a dashed vertical line. Their parameters (mean and standard deviation) are also indicated in each panel.}
    \label{fig:L6_Lx_dif_relations}
\end{figure}

Figure \ref{fig:L6_Lx_dif_relations} shows the offsets distribution between the measured $L_X$ and the predicted X-ray luminosity for each of the 3 relations listed above.
The offset is the difference between the median of the log$L_X(\rm 2-10\,keV)$ posterior distribution and the predicted logarithmic luminosity by each relation using the inferred $L_{6\mu m}$ of the sources. The $\Delta L_X$ distributions are fit with a Gaussian to infer the corresponding mean and scatter. The best-fit parameters are also shown in Figure \ref{fig:L6_Lx_dif_relations}. The relations of \cite{Chen_2017} and \cite{Stern} show a narrower dispersion of the $\Delta L_X$ distributions than the one by \cite{Mateos}. Besides, our observations show systematic offsets relative to the \cite{Chen_2017} or \cite{Mateos} relations. This is smaller in the case of the \cite{Stern} relation. 
We have confirmed that fitting a second order polynomial to the data points shown in Figure \ref{fig:L6_Lx_plot} yields a $L_{6\mu m}-L_X(\rm 2-10\,keV)$ relation similar to that of \cite{Stern}. Since the AGN sample used in \cite{Stern} is larger than ours, spans a broader luminosity baseline and is independently selected, we choose to use their $L_{6\mu m}-L_X(\rm 2-10\,keV)$ correlation in our analysis. Our final results and conclusions are not sensitive to that choice. 

For the SED constrained sources we convert the $L_{6\mu m}$ luminosity PDFs into a $L_X(\rm 2-10\,keV)$ PDF by using the \cite{Stern} relation. We account for the dispersion of this relation by convolving the inferred PDFs with a Gaussian with a 0.4 dex logarithmic standard deviation. The latter value is the dispersion estimated by \cite{Stern} and is similar to the standard deviation of our $\Delta L_X$ distribution in Figure \ref{fig:L6_Lx_dif_relations}.
In the case of $6\mu m$ AGN luminosity upper limits, we assume that all luminosities below the $3\sigma$ upper limits are equiprobable. For this reason, we consider the $L_{6\mu m}$ PDFs to be log-uniform below the upper limits and have a zero-probability for brighter luminosities.
We also apply the \cite{Stern} relation on it and convolve it with the best fit Gaussian to obtain the $L_X$ prior.

The X-ray luminosity is not one of the free parameters of the UXCLUMPY model. Therefore, we translate the X-ray luminosity prior to the UXCLUMPY normalisation parameter prior, which is then applied to the BXA spectral fits. We can then reprocess the X-ray spectra adding the mid-IR information for the selected X-ray AGN. This prior aims to improve the characterisation of highly obscured AGN and only minimally disrupt the X-ray spectral fits. We, therefore, choose to apply it only to potential CTK sources, defined as those with at least 5\% of their posterior $N_{\rm H}$ distribution being above the CTK limit, $10^{24}$cm$^{-2}$. This definition is the same as in \citetalias{Lanzuisi_2018}. Changing this cut to 1\%  or 10\% has little impact on the final results. There are 829 CTK candidates among our sources, but 54 of them do not have an IR-counterpart in \cite{Jin_2018} catalogue. We then apply our methodology to a sample comprising 775 sources, of which 314 have constrained priors and 461 have upper limits priors.

\begin{figure}
    \centering
    \includegraphics[width=0.45\textwidth]{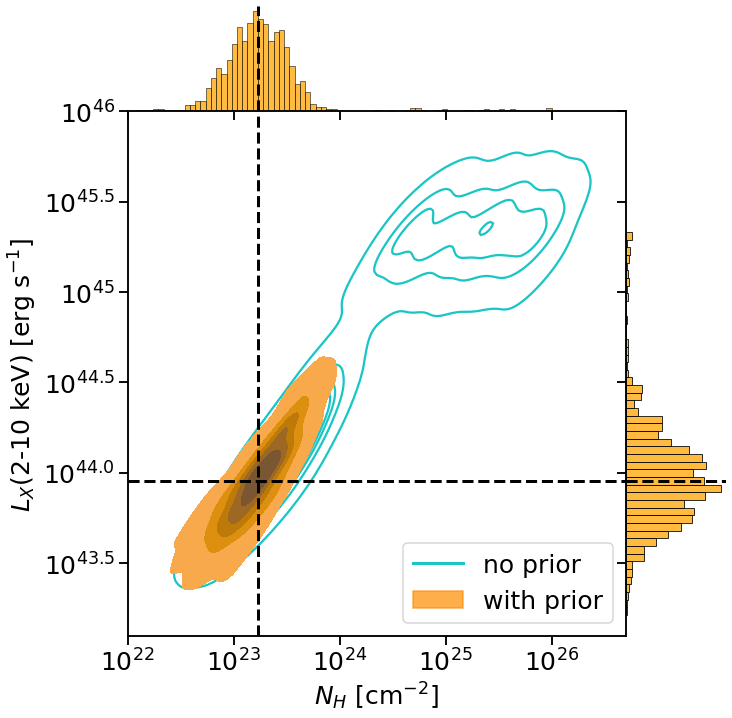}
    \caption{Density plot of the posterior distribution of the X-ray luminosity $\log L_X(\rm 2-10\,keV)$ as a function of the column density $\log N_{\rm H}$ for the source shown in Figures \ref{fig:double_peaks_Lx_NH_scatter} and \ref{fig:IR_SED}. The parameter posterior distributions are obtained by X-ray spectral analysis using the $L_{6\mu m}$-based luminosity prior. The 1-dimensional projections of the posterior on the X-ray luminosity and column density axes are shown on the right and top panels, respectively. The median values are plotted in dashed black lines. The posterior distribution of the fit without the use of prior is plotted as cyan contours in the background.}
    \label{fig:double_peak_scatter_after_prior}
\end{figure}

For the same source shown in Figures \ref{fig:double_peaks_Lx_NH_scatter} and \ref{fig:IR_SED},  Figure \ref{fig:double_peak_scatter_after_prior} demonstrates how the  L$_{6\mu m}$-based prior breaks the X-ray spectral modelling degeneracies and improves the column density constraints. 
Before applying the $L_{6\mu m}$-based prior, there is a total of 300 sources with broad or multi-modal X-ray analysis posteriors and IR counterparts.  After, only 16 of them remained with  multi-modal/broad posteriors. 15 of these sources have less than 35 photons in their X-ray spectra. The lack of certainty in these cases is not surprising, but our methodology nonetheless helps constrain the physical parameters of the bulk of the population.

\begin{figure}
    \centering
    \includegraphics[trim = 0cm 0cm 0cm 0cm, clip,width=0.45\textwidth]{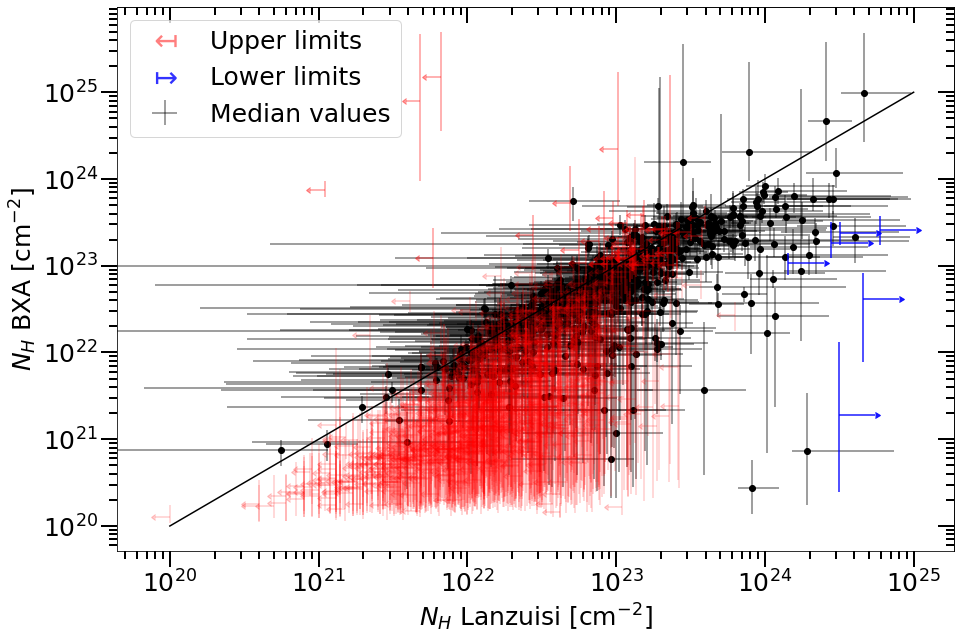}
    \caption{Comparison of the $N_{\rm H}$ values derived from  our spectral analysis using the $L_{6\mu m}$ prior (vertical axis) with those derived by \citetalias{Lanzuisi_2018} (horizontal axis). The black dots represent the median values of our spectral fit estimations. The red (blue) arrows are upper (lower) limits in \citetalias{Lanzuisi_2018}.}
    \label{fig:compa_marchesi_w_prior}
\end{figure}

In Figure \ref{fig:compa_marchesi_w_prior}, the $N_{\rm H}$ values obtained from the fit using priors are plotted against the values of \citetalias{Lanzuisi_2018}. The figure is similar to figure \ref{fig:compa_Marchesi} as we only apply the $L_{6\mu m}$-based prior on the potentially CTK sources. There are nonetheless significant differences because our updated spectral fits now yield a smaller number of CTK sources compared to \citetalias{Lanzuisi_2018}. This is because the CTK sources in \citetalias{Lanzuisi_2018} deviate from the $L_X-L_{6\mu m}$ correlations \citep[e.g.][]{Stern} in the sense that they are systematically overluminous at X-rays for their $6\mu m$ luminosity. This point has been acknowledged by \citetalias{Lanzuisi_2018}, where their CTK sources are systematically offset by more than 1$\sigma$ from the \cite{Stern} relationship (see Figure 4 in \citetalias{Lanzuisi_2018}). 

\begin{figure}
    \centering
    \includegraphics[trim = 0cm 0cm 0cm 0cm, clip,width=0.48\textwidth]{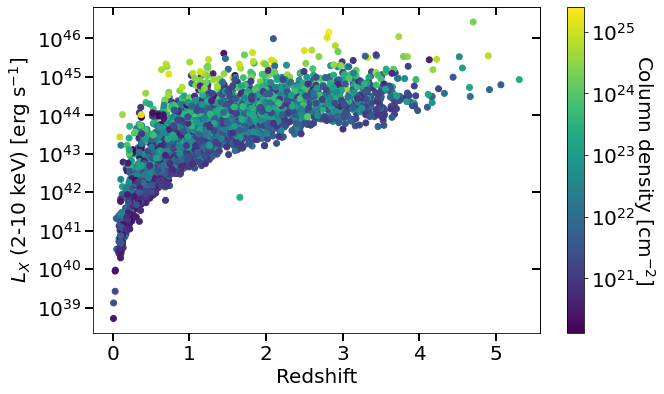}
    \caption{Intrinsic X-ray luminosity (2-10 keV) against the redshift for the sources of our sample. The colour indicates the column density of the source in cm$^{-2}$. The values used in this figure correspond to the median of the posterior distribution of the respective parameters. The points' locations are not very representative in the case of broad distributions, for example, in the case of the outlier source at $z\sim 1.7$ with $L_X\sim 10^{42}\rm \, erg \, s^{-1}$.}
    \label{fig:Lx_vs_z}
\end{figure}

Table \ref{tab:fit_information} is an extract of the table summarizing the results of our sample's X-ray spectral analysis after using the $L_{6\mu m}$-based prior. The uncertainties of the principal parameters are also indicated. For an overall summary of our sample parameter estimations, Figure \ref{fig:Lx_vs_z} displays the intrinsic X-ray luminosity (2-10 keV) as a function of the redshift and coloured as a function of the median column density.

\begin{table*}
    \centering
    \caption{X-ray spectral fitting results. (1) source ID; (2) redshift; (3) redshift type: spectroscopic, photometric or None; (4) X-ray 2-10keV logarithmic luminosity and its 1-$\sigma$ uncertainty; (5) logarithmic column density $N_{\rm H}$ and its 1-$\sigma$ uncertainty; (6) photon index $\Gamma$ and its 1-$\sigma$ uncertainty; (7) CTK candidate flag i.e. if the original spectroscopic fit includes more than 5\% of its posterior distribution in the CTK regime; (8) double-peaked flag (see definition in section \ref{compaLanzuisi_subsec}); (9) source ID in the multiwavelength catalog \citep{Jin_2018} if available;
    (10) logarithmic AGN $L_{6\mu m}$ from SED fitting and its 1-$\sigma$ uncertainty; (11) logarithmic AGN $L_{6\mu m}$ upper limit at 99 percentile if the SED fit is unconstrained. Full table electronically available.}

    \begin{tabular}{c|c|c|c|c|c|c|c|c|c|c}
    ID & z & ztype & log($L_X$) & log($N_{\rm H}$) & $\Gamma$ & CTK  & 2-peaked & ID  & log($L_{6\mu m}$) & $L_{6\mu m}$ upp. lim\\
    &&&[erg s$^{-1}$]&[cm$^{-2}$]&&candidate&&\cite{Jin_2018}&[erg s$^{-1}$] & [erg s$^{-1}$]\\ 
    (1) & (2) & (3) & (4) & (5) & (6) & (7) & (8) & (9) & (10) & (11) \\ \hline
    COSMOS\_0\_1& 1.342& zphot&        $44.25^{+0.08}_{-0.11}$&        $21.10^{+0.76}_{-0.76}$&        $1.90^{+0.13}_{-0.13}$&        False & False& --  &     -- & --\\
    COSMOS\_0\_10& 1.283& zphot&        $43.52^{+0.27}_{-0.25}$&        $22.38^{+0.36}_{-0.82}$&        $1.94^{+0.15}_{-0.14}$&        False & False&  10043855 &     -- & 43.82\\
    COSMOS\_0\_100& 0.582& zphot&        $43.01^{+1.18}_{-0.70}$&        $22.16^{+0.62}_{-0.31}$&        $1.96^{+0.13}_{-0.15}$&        False & False&  10050161 &     $42.56^{+0.27}_{-0.19}$& --\\
    COSMOS\_0\_101& 0.619& zphot&        $42.98^{+0.12}_{-0.14}$&        $22.41^{+0.15}_{-0.19}$&        $1.93^{+0.14}_{-0.15}$&        False & False& 10051883  &     -- & 43.14\\
    COSMOS\_0\_102& 0.516& zphot&        $42.33^{+0.73}_{-1.17}$&        $21.35^{+0.62}_{-0.90}$&        $1.93^{+0.14}_{-0.16}$&        False & False&  -- &     -- & --\\
    \vdots&\vdots&\vdots&\vdots&\vdots&\vdots&\vdots&\vdots&\vdots&\vdots&\vdots\\
    COSMOS\_8\_95& 2.212& zphot&        $44.01^{+0.37}_{-0.93}$&        $21.72^{+0.76}_{-1.01}$&        $1.87^{+0.14}_{-0.13}$&        False & False&  10203746 &     -- & 44.29\\
    COSMOS\_8\_96& 0.205& zphot&        $41.72^{+0.09}_{-0.10}$&        $21.37^{+0.35}_{-0.56}$&        $1.84^{+0.14}_{-0.13}$&        True & True&   10203914  &   $42.34^{+0.10}_{-0.09}$& --\\
    COSMOS\_8\_97& 1.608& zspec&        $44.43^{+0.04}_{-0.05}$&        $21.57^{+0.42}_{-0.85}$&        $1.96^{+0.11}_{-0.11}$&        False & False&  10204309 &     $44.63^{+0.18}_{-0.17}$& --\\
    COSMOS\_8\_98& 2.620& zspec&        $44.56^{+0.11}_{-0.09}$&        $22.00^{+0.67}_{-1.26}$&        $1.87^{+0.14}_{-0.14}$&        False & False&  -- &     -- & --\\
    COSMOS\_8\_99& 0.686& zphot&        $42.96^{+0.11}_{-0.15}$&        $20.68^{+0.56}_{-0.48}$&        $2.05^{+0.13}_{-0.15}$&        False & False&  20010065 &     $43.51^{+0.09}_{-0.08}$& --\\

    \end{tabular}
    
    \label{tab:fit_information}
\end{table*}

\section{Obscured AGN demographics}\label{Obscured_demo_sec}


    \subsection{Observed parameter distribution}\label{obs_param_distrib_subsec}

\begin{figure}
    \centering
    \includegraphics[width=0.45\textwidth]{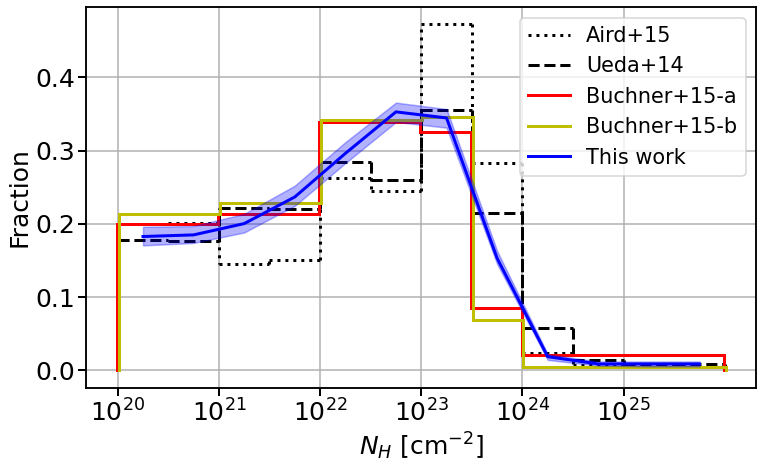}
    \caption{$N_{\rm H}$ distribution of the hard band detected sources in the \textit{Chandra} COSMOS Legacy field. The thick blue line corresponds to the constraints from our X-ray spectral analysis. It represents the median of the bootstrap resampling approach described in the text. The light blue shaded region corresponds to the 68\% confidence interval around the median. The black dotted line and the black dashed histograms show the predicted $N_{\rm H}$ distributions obtained using the \protect\cite{Aird_2015} and \protect\cite{Ueda_2014} luminosity functions, respectively. The red and yellow lines represent the redshift distribution estimated in \protect\cite{Buchner_2015} obtained by using a constant-value prior and a constant-slope prior, respectively.}
    \label{fig:distribution_NH_vs_prediction}
\end{figure}

One of the motivations of the analysis presented in this work is to place constraints on the demographics of obscured AGN. Therefore, in the next sections, we focus on the hard-band (2-7\,keV) selected sample of the \textit{Chandra} COSMOS Legacy survey. This is because photons at rest-frame energies $>2$\,keV can penetrate relatively dense columns of gas clouds, thereby providing a better handle on the obscured AGN population.  Figure \ref{fig:distribution_NH_vs_prediction} displays the LOS column density $N_{\rm H}$ distribution for all sources selected in the hard X-ray band.
The histogram in this figure is constructed from the $N_{\rm H}$ posterior distributions derived by fitting the X-ray spectra of individual COSMOS sources with the baseline X-ray spectral model described in section \ref{uxclumpy_subsection} and with the $L_{6\mu m}$-based prior (i.e. Section \ref{L6improve_subsec}). A bootstrap resampling approach is adopted, whereby the posterior distribution of each X-ray source is resampled with replacement to generate 100 realisations of the population. These are then used to determine, within the different $N_{\rm H}$-bins, the median fraction and 16th and 84th percentiles, corresponding to the 1$\sigma$ variation lower and upper limits. Figure \ref{fig:distribution_NH_vs_prediction} shows that the our sample includes a large observed fraction of obscured AGN ($59.7^{+0.2}_{-0.7}$\% ) with LOS column densities $\rm N_{\rm H}>10^{22}\,cm^{-2}$. However, the sensitivity of this survey drops close to and above the CTK limit. This is evident from the decreasing fraction of AGN in Figure \ref{fig:distribution_NH_vs_prediction} toward column densities $\rm N_{\rm H}\approx\rm 10^{24}\,cm^{-2}$.

\begin{figure}
    \centering
    \includegraphics[width=0.5\textwidth]{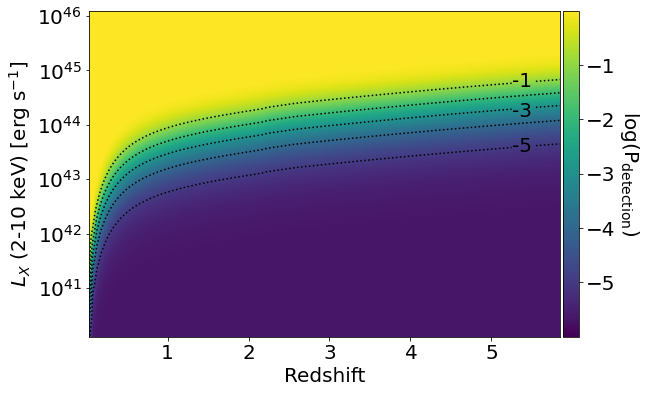}
    \caption{Sensitivity map showing the probability of detection of a CTK source with log($N_{\rm H}$) $=\rm 24.1\,cm^{-2}$ within \textit{Chandra} COSMOS Legacy as a function of its intrinsic X-ray luminosity and redshift. The probability is given on a logarithmic scale. The black dotted lines indicate detection probabilities from $10^{-1}$ (top) to $10^{-5}$ (bottom) in logarithmic steps of 1\,dex. As expected, the detection probability rapidly decreases with the increasing redshift and increases with the luminosity.}
    \label{fig:sensitivity_map}
\end{figure}

We also over-plot in Figure \ref{fig:distribution_NH_vs_prediction} the predictions from X-ray Luminosity Function (XLF) models in the literature,  \cite{Ueda_2014}, \cite{Aird_2015} and \cite{Buchner_2015}. 
The XLFs encapsulate the intrinsic number of sources as a function of their physical properties (obscuration, X-ray luminosity and redshift). To convert these intrinsic source numbers into observed source numbers, we need to convolve them with the \textit{Chandra} sensitivity maps.
To calculate the sensitivity maps, we use the UXCLUMPY model to predict the expected \textit{Chandra}/ACIS photon rate for an AGN of a given redshift, 2-10\,keV luminosity and absorbing hydrogen column density. For the calculation of photon rates, we assume a Gaussian photon index distribution (mean 1.9, scatter 0.15) for UXCLUMPY and a soft-excess component logarithmic normalisation that is uniformly distributed in the range $10^{-7}-10^{-1.5}$. We marginalise over these two parameters to calculate the average photon rate $\nu(z, L_X, N_{\rm H})$. This can then be converted into the area over which such a source can be detected by using the sensitivity curves derived in section \ref{COSMOS_subsec}. The end products of these calculations are cubes describing the survey area available to sources as a function of $L_X$, z, $N_{\rm H}$. By normalising them, one obtains the detection probability of a source as a function of its intrinsic characteristics. Figure \ref{fig:sensitivity_map} shows the 2-dimensional projection of such a cube on the redshift-luminosity plane for a CTK source with $N_{\rm H}=\rm 1.26 \times 10^{24}\,cm^{-2}$. This figure shows that the probability of detecting such a source decreases toward lower luminosities and higher redshift. The analytic XLFs are convolved with the sensitivity maps and then integrated over luminosity and redshift to yield the predictions on the number of AGN as a function of $N_{\rm H}$ in Figure \ref{fig:distribution_NH_vs_prediction}.

The comparison with predictions is intended to guide the expected $\rm N_{\rm H}$ distribution of AGN in the COSMOS field based on established knowledge of their demographics. We caution that comparing these model predictions and the observations should be taken with a grain of salt. This is because the histogram of the posteriors in Figure \ref{fig:distribution_NH_vs_prediction} represents the convolution of the intrinsic column density distribution of AGN with the observational uncertainties. Instead, the model XLFs do not include such errors. It is nevertheless interesting that the overall shape of the model/observed column density distributions is similar. There is an increase in the number of sources with increasing column density to $\log N_{\rm H}\rm / cm^{-2}\approx23.5$ followed by a steep decline for higher levels of obscuration.

\begin{figure}
    \centering
    \includegraphics[width=0.47\textwidth]{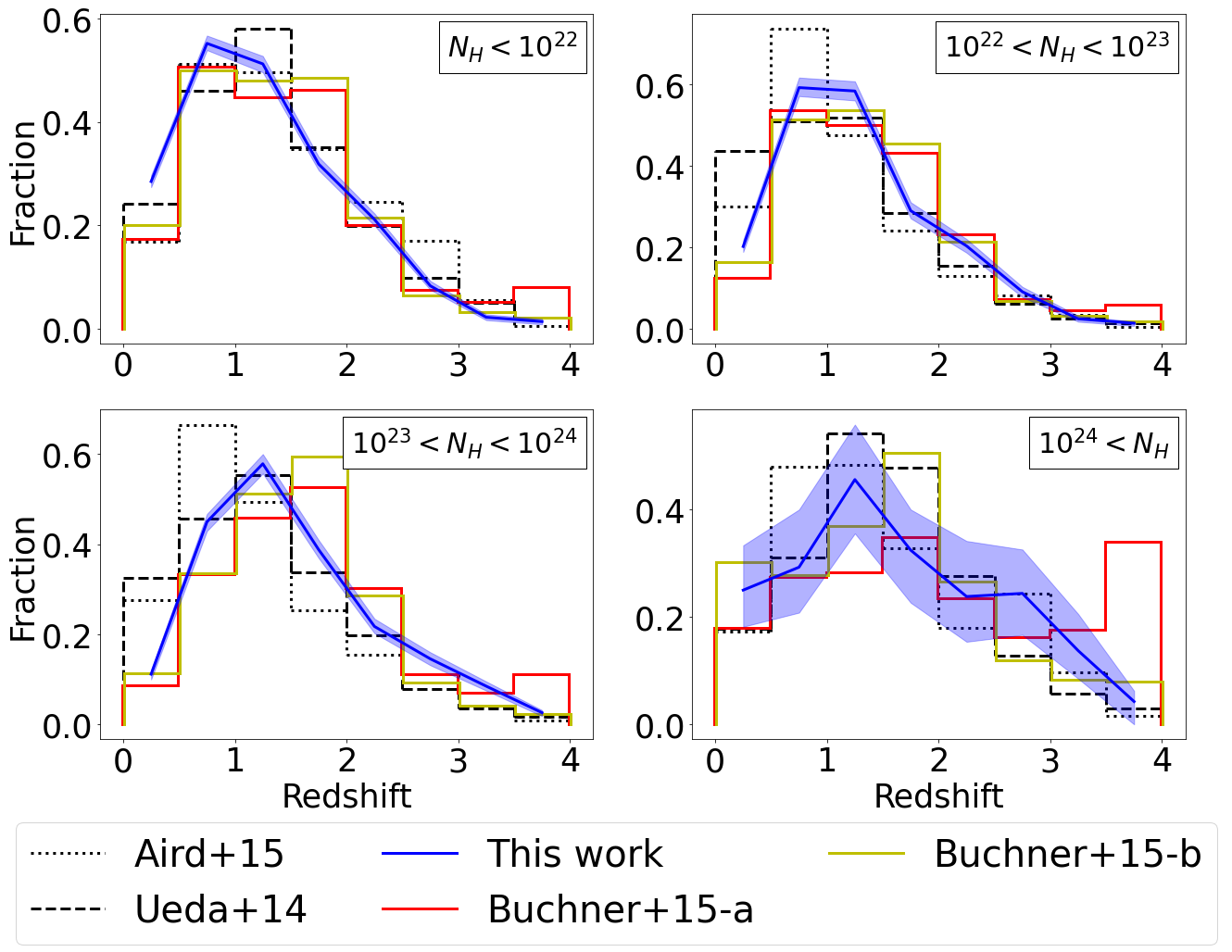}
    \caption{Redshift distributions of the hard band detected sources in the \textit{Chandra} COSMOS Legacy field. Each panel corresponds to a different $N_{\rm H}$ interval. The thick blue line is the median of the bootstrap resampling approach described in the text. The light blue shaded region corresponds to the 68\% confidence interval around the median. When available, the redshift information comes from spectroscopy or otherwise from the redshift posterior distributions generated by the X-ray spectral fitting analysis using the photometric redshift distribution as a prior. The black dotted line and the black dashed histograms show the predicted $N_{\rm H}$ distributions obtained using the \protect\cite{Aird_2015} and \protect\cite{Ueda_2014} luminosity functions, respectively. The red and yellow lines represent the redshift distribution estimated in \protect\cite{Buchner_2015} obtained by using a constant-value prior and a constant-slope prior, respectively.}
    \label{fig:distribution_z_vs_prediction}
\end{figure}

Figure \ref{fig:distribution_z_vs_prediction} further explores the redshift distribution of the AGN for different $N_{\rm H}$ intervals. In this plot, our histograms are constructed using the same bootstrapping methodology described above. Both the model and the observed distributions in Figure \ref{fig:distribution_z_vs_prediction} show a broad peak at $z\approx1$ and a decline to higher redshift. This behaviour can be explained by the overall redshift evolution of the X-ray luminosity function and the flux limit of the COSMOS Legacy survey, which yields increasingly smaller AGN samples at higher redshift. This trend is broadly reproduced by the XLFs of \cite{Ueda_2014}, \cite{Aird_2015} and \cite{Buchner_2015}.

    \subsection{Space density measurements}\label{space_density_subsec}

This section describes how the X-ray spectral analysis results are combined with the X-ray selection function of the COSMOS Legacy survey for the hard band-selected sources, to determine the space density of AGN as a function of the redshift $z$, X-ray luminosity $L_X(\rm 2-10\,keV)$, and column density $N_{H}$.

Given a model of AGN space density, $\phi(L_X, z, N_{\rm H})$, described by a set of parameters, $\Psi$, the likelihood of a given a set of observations, $D$, is described by the product of the Poisson probabilities of individual sources:
\begin{multline}
    \mathcal{L}(D | \Psi) = e^{-\lambda} \prod_i^n \int d\log L_X \int d\log N_{\rm H} \int \frac{dV}{dz} dz     \\ 
    p(z, L_X, N_{\rm H} | D_i) \phi(L_X, z, N_{\rm H} | \Psi) \, ,
    \label{Likelihood}
\end{multline}
\noindent where $n$ is the number of individual sources in the field, $i$ is their index, $\frac{dV}{dz}$ is the differential co-moving volume. Then, $p(z, L_X, N_{\rm H} | D_i)$ is the probability that a source has a luminosity $L_X$, a redshift $z$, and column density $N_{\rm H}$. It encapsulates the uncertainty of deriving these values from the observations, i.e. from X-ray spectral analysis or in the case of photometric redshifts from the multi-waveband SED fits. The quantity $\lambda$ is the expected number of AGN in the survey as a function of the parameters set and is defined as 

 \begin{equation}
    \lambda = \int d\log L_X \int d\log N_{\rm H} \int \frac{dV}{dz} dz \, A(z, L_X, N_{\rm H}) \phi(L_X, z, N_{\rm H} | \Psi),
    \label{lambda}
\end{equation}

\noindent  where $A(z, L_X, N_{\rm H})$ is the sensitivity curve representing the area of the survey for which a source with $z$, $L_X$ and $N_{\rm H}$ can be detected and its calculation is described in section \ref{obs_param_distrib_subsec}.

We decide to use a non-parametric approach to determine the luminosity function \citep[following][]{Buchner_2015, Georgakakis_2017}. The $z$, $L_X$, $N_{\rm H}$ parameter space is divided into a 3-dimensional grid, and in each grid cell, the luminosity function is assumed to be constant. 
Using such a non-parametric approach allows the space density to vary more freely and eventually find large variations of shape across the parameter grid. 
The edges of the grid pixels in each of the three dimensions are log($L_X$)=(40.0, 41.0, 42.0, 42.5, 43.0, 43.5, 44.0, 44.5, 45.0, 46.0, 47.0) [log(erg s$^{-1}$)], $z$=(0.0, 0.5, 1.0, 1.5, 2.0, 2.5, 3.0, 6.0) and log($N_{\rm H}$)=(20.0, 22.0, 23.0, 24.0, 26.0) [log(cm$^{-2}$)]. The total number of free parameters is 280. The likelihood (Equation \ref{Likelihood}) is integrated using the principle of the Importance sampling \citep{press}. 
It is also worth emphasising that in the Bayesian framework of Equation  \ref{Likelihood} the posterior distribution  $p(z, L_X, N_{\rm H} | D_i)$ of a given source (i.e. contours and shaded regions in Figures  \ref{fig:double_peaks_Lx_NH_scatter} and  \ref{fig:double_peak_scatter_after_prior}) is weighted by the luminosity function  $\phi(L_X, z, N_{\rm H})$ when estimating the likelihood. Non-physical posterior solutions, e.g. very high AGN luminosities, can therefore be weighted down a posteriori because they are rare. This is the case of parametric XLF studies that typically assume a double power-law form with a relatively steep bright-end slope. In our non-parametric approach, however, there is no imposed shape, and the AGN space density at each grid point is independently determined. In that respect, our analysis is more sensitive to broad $p(z, L_X, N_{\rm H} | D_i)$ posteriors like those shown in Figure  \ref{fig:double_peaks_Lx_NH_scatter}. We nevertheless compensate for that using the multiwavelength priors described in Section \ref{L6improve_subsec} to narrow down the X-ray spectral analysis posteriors of the sample sources.

\begin{figure*}
    \centering
    \includegraphics[width=\textwidth]{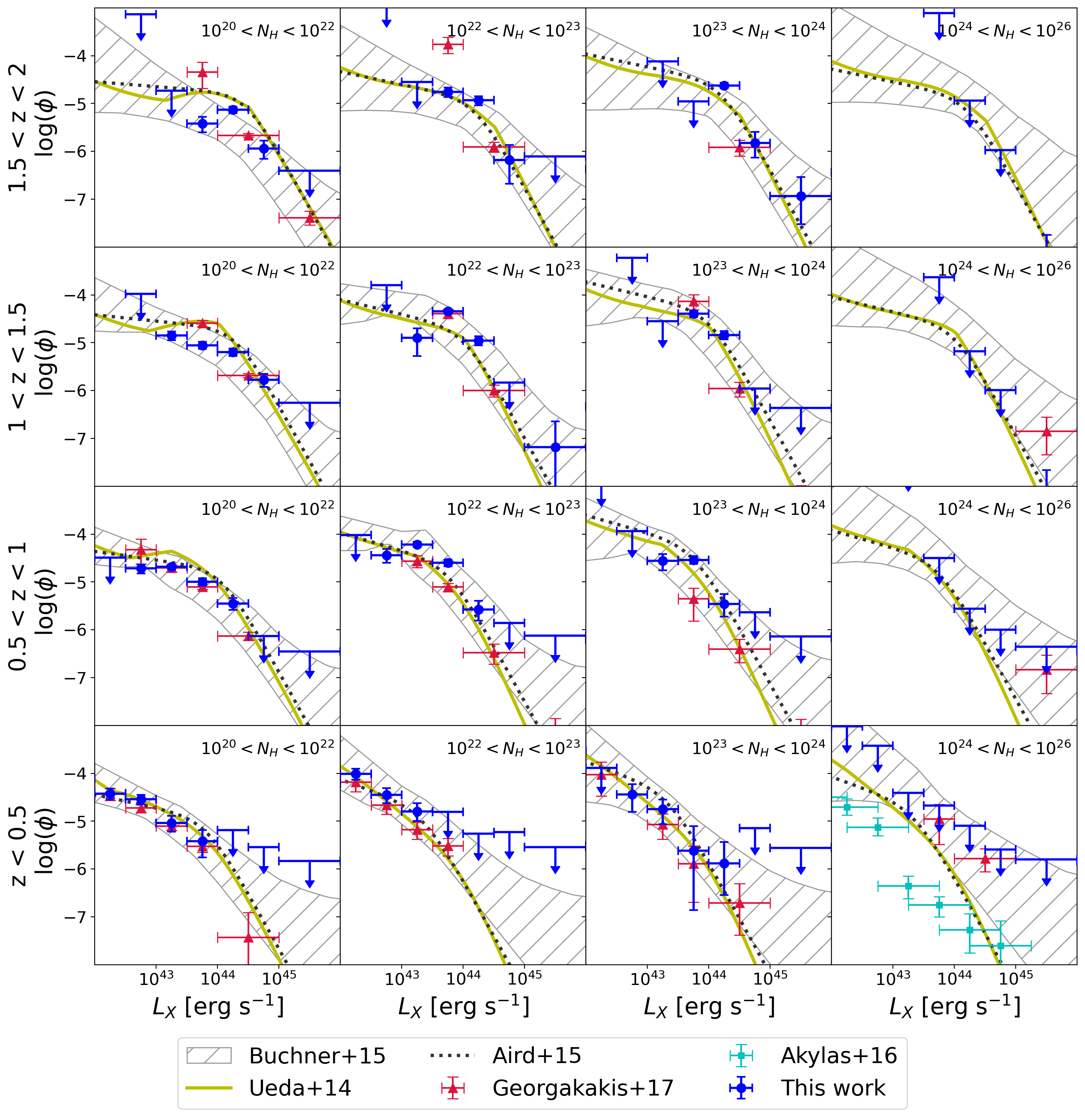}
    \caption{Space density curves ($\rm Mpc^{-3}\,dex^{-2}$) as a function of the intrinsic X-ray luminosity in the 2-10\,keV band for different redshift and column density intervals. 
    The redshift range of each panel row is indicated on its left side. From top to bottom, the rows correspond to $z=1.5-2$, $z=1-1.5$, $z=0.5-1$ and $z=0-0.5$.
    The column density increases from left to right, as indicated at the bottom of each column. The constraints from our analysis are shown with the blue dots and associated to 1$\sigma$ uncertainties and, in the case of the upper limits, with the blue arrows pointing down. A space density is considered an upper limit if its 1$\sigma$ uncertainty width is larger than 2 dex.
    These upper limits correspond to the $3\sigma$ confidence interval. The yellow solid line and the black dotted line correspond to \protect\cite{Ueda_2014} and \protect\cite{Aird_2015} luminosity functions, respectively. The grey hatched area represents the non-parametric constraints of \protect\cite{Buchner_2015}. The red triangles are the luminosity function measured by \protect\cite{Georgakakis_2017} in the XMM-XXL field \protect\citep{Pierre2016}. The CTK luminosity function determined by \protect\cite{Akylas_2016} in the local Universe is shown with the cyan squares.}
    \label{fig:space_density}
\end{figure*}

We use STAN, a Hamiltonian Markov chain Monte Carlo code \citep{STAN} to sample the likelihood (Eq. \ref{Likelihood}) in a Bayesian framework and to obtain the space density posterior distribution for each cell of the 3-dimensional parameter grid.
The resulting space density measurements are displayed in Figure \ref{fig:space_density} and compared with different X-ray luminosity functions in the literature. We limit this comparison to $z<2$, where the COSMOS Legacy survey provides strong constraints. At higher redshift, the number of X-ray AGN in our sample decreases rapidly, and our space density measurements suffer larger uncertainties.

Our space density measurements have a broad overall agreement with previous studies. In detail, however, there are subtle differences. For example, in the case of unobscured AGN, $N_{\rm H}\rm <10^{22}\, cm^{-2}$, our constraints lie systematically lower than the analytic XLFs of \cite{Ueda_2014} and \cite{Aird_2015}, particularly for luminosities $L_X(\rm 2-10\,keV)\approx10^{43}-10^{44} \, erg \, s^{-1}$. This trend is stronger for the redshift intervals $z=0.5-1$ and $1-1.5$  but is also evident in other redshift bins. For moderately obscured AGN in the range $N_{\rm H}=\rm 10^{22}-10^{23}\,cm^{-2}$, our measurements are instead systematically higher than those of  the analytic XLFs of \cite{Ueda_2014} and \cite{Aird_2015}. These differences are partly related to the fact that the NH constraints in these studies are largely based on hardness ratios, which have limited discriminating power for AGN close to the $10^{22}$cm$^{-2}$ limit. It is nevertheless reassuring that for the integrated space densities in the interval $N_{\rm H} = 10^{20} - 10^{23}$ cm$^{-2}$,  the agreement between our analysis and the analytic studies above is good. In any case, these differences also highlight the importance of fully non-parametric XLF approaches, like the one presented here, to supplement and guide analytic prescriptions. In the CTK regime, our analysis only yields upper limits in the AGN space density because of the low number of CTK sources in our sample. Nevertheless, the $3\sigma$ upper limits, particularly at $z>0.5$, favour CTK space densities at the low-end of the range covered by the previous observational constraints shown in Figure \ref{fig:space_density}.

    \subsection{Compton-Thick fraction}\label{CT_fraction_results_subsec}

The intrinsic CTK fraction, $f_{CTK}$, is an important parameter for characterizing the demographics of heavily obscured AGN. In our analysis, this is defined as the ratio between the number of AGN with column density $N_{\rm H}=\rm 10^{24}-10^{26}\,cm^{-2}$ and those with $N_{\rm H}=\rm 10^{20}-10^{26}\,cm^{-2}$:

\begin{equation}
    f_{CTK} = \frac{N_{24-26}}{N_{20-22} + N_{22-23} + N_{23-24} + N_{24-26}}.
    \label{eq_fCT}
\end{equation}

\noindent In the equation above, $N$ refers to the intrinsic number of AGN at different logarithmic column density intervals. For the estimation of  $f_{CTK}$, we slightly modify the X-ray luminosity function model used in the Bayesian inference methodology. The space density of CTK AGN at all redshift and luminosity intervals is linked to that of CTN AGN via Equation \ref{eq_fCT} where $f_{CTK}$ is a free parameter of the model. After some algebra, the CTK space density is estimated as 

\begin{equation}
\begin{split}
    \phi(L_X, z, \log N_{\rm H}=24-26) & =  \frac{f_{CTK}}{2\cdot(1 - f_{CTK})}\,\cdot\\ 
    &  \bigg(  2\cdot\phi(L_X, z, \log N_{\rm H}=20-22) + \\
    & \phi(L_X, z, \log N_{\rm H}=22-23) + \\
    & \phi(L_X, z, \log N_{\rm H}=23-24) \bigg). \\
    \label{eq_fCT_STAN}
\end{split}
\end{equation}

\noindent The multiplicative factors of 2 in the above equation account for broader logarithmic column density intervals for CTK ($\log N_{\rm H}/\rm cm^{-2}=24-26$) and unobscured ($\log N_{\rm H}/\rm cm^{-2}=20-22$) AGN.  The $f_{CTK}$ is assumed here to depend only on the redshift but not on the luminosity. This is equivalent to assuming that at fixed redshift, the CTK luminosity function is related to the CTN one via a scaling factor \citep[e.g.][]{Aird_2015}. 
The $f_{CTK}$ parameter is therefore independently determined for each of the redshift intervals adopted in section \ref{space_density_subsec} with edges $z=$ (0.0, 0.5, 1.0, 1.5, 2.0, 2.5, 3.0, 6.0). The Stan Hamiltonian Markov chain Monte Carlo code then yields a posterior distribution of the $f_{CTK}$ for each of the redshift intervals above.

\begin{figure}
    \centering
    \includegraphics[width=0.5\textwidth]{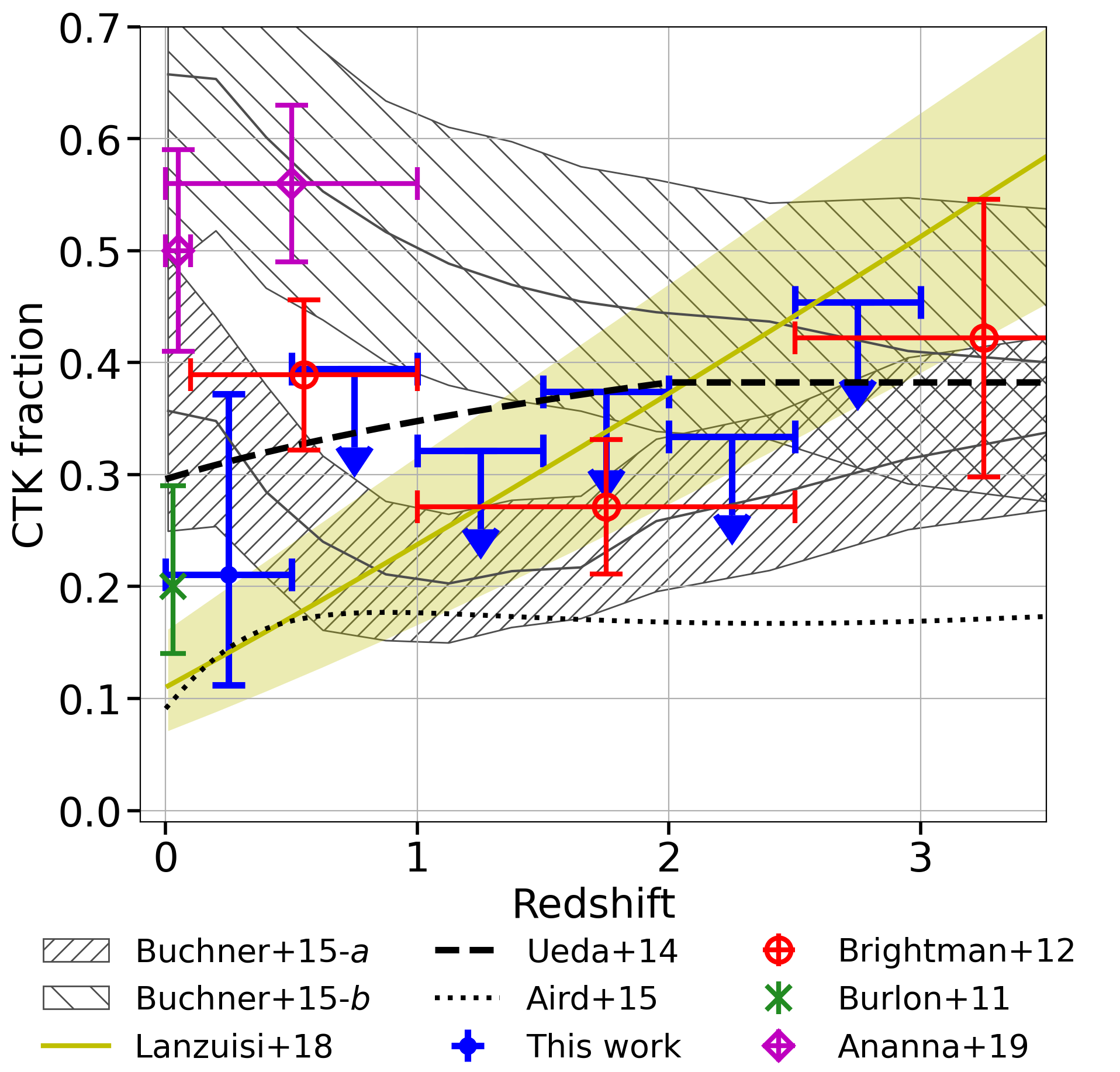}
    \caption{Intrinsic CTK fraction as a function of redshift. The blue symbols show the results of our analysis. The constraints for the redshift interval $z=0-0.5$ (blue circle) show the mode of the posterior distribution and the associated 68\% confidence interval. At higher redshift intervals, the downward blue arrows correspond to the 3$\sigma$ upper limit estimated from the corresponding posterior distributions. The horizontal error bars of all blue symbols show the width of the redshift intervals. The vertically (Buchner+15-a) and horizontally (Buchner+15-b) hatched shaded regions represent the 1$\sigma$ uncertainty of the CTK fractions presented in \protect\cite{Buchner_2015} obtained by using a constant-value prior and a constant-slope prior, respectively. The solid line within these regions corresponds to the median. The CTK fraction of the  \protect\cite{Ueda_2014} and \protect\cite{Aird_2015} X-ray luminosity functions are shown with the black dashed and the black dotted lines, respectively. The green cross at $z\approx0$ is the measured CTK fraction in the local Universe determined by \protect\cite{Burlon_2011}. The red circles are the results of \protect\cite{Brightman_2012}. The yellow shaded region is the CTK fraction estimation by \protect\citetalias{Lanzuisi_2018}. The pink diamonds correspond to the CTK fraction estimated by \protect\cite{Ananna2019} at redshift z<0.1 and z<1.0.}
\label{fig:CT_fraction}
\end{figure}

Figure \ref{fig:CT_fraction} plots the $f_{CTK}$ inferred from our analysis as a function of redshift for the hard band selected AGN. This parameter is constrained to be $21.0^{+16.1}_{-9.9}\%$
($1\sigma$) for the lowest redshift interval $z=0-0.5$. At higher redshifts, only upper limits to $f_{CTK}$ can be derived. Also shown in Figure \ref{fig:CT_fraction} are previous results on the CTK fraction that demonstrate the range of values covered by different studies. At $z=0-0.5$, our analysis favours CTK fractions at the low end of the distribution of $f_{CTK}$ in the literature. At higher redshift, our $3\sigma$ upper limits are generous, but at least for $z\la2$, suggest $f_{CTK}\la40\%$. 

\section{Discussion}\label{Discussion_sec}

    \subsection{Obscuration characterisation}
A multiwavelength Bayesian approach is presented that combines X-ray spectral fits with information from mid-IR wavelengths to constrain the level of LOS obscuration of AGN. This is motivated by the need to resolve degeneracies between two key physical parameters inferred from X-ray spectral analysis, the intrinsic AGN luminosity and the LOS hydrogen column density. These covariances are particularly important for sources with a low number of counts and high levels of obscurations leading to broad or multi-modal posterior distributions. Our methodology addresses this issue by using the mid-IR part of the SED as a prior for the intrinsic AGN luminosity. This is based on the expectation that the short wavelength (X-rays, UV) AGN radiation absorbed by dust and gas clouds emerges as thermal emission in the IR. Support for such an energy balance between different parts of the SED is coming from observed correlations between the intrinsic (i.e. corrected for obscuration) X-ray luminosity and the mid-IR luminosity of AGN  \citep{Gandhi_2009, Stern}. These correlations are claimed to apply to AGN over a broad range of LOS obscurations, including CTK sources \citep{Gandhi_2009, Asmus2015, Annuar2017}. Therefore, our methodology relies on such relations as an independent handle on the intrinsic accretion luminosity, particularly in the case of obscured systems. We caution, however, that in the local Universe, there are (few) examples of heavily obscured (CTK) AGN that appear subdominant in the mid-IR for their intrinsic X-ray luminosity \citep{Krabbe_2001, Gandhi_2015} and, therefore, deviate from the established correlations. For these sources, the mid-IR photons may be absorbed by the obscuring medium. Our methodology is not optimal for this class of AGN.

We apply our approach to X-ray sources in the \textit{Chandra} COSMOS Legacy survey to constrain in a non-parametric way the space density of AGN in bins of luminosity, redshift and column density. We find small systematic differences between our results and previous parametric estimates of the AGN X-ray luminosity function in the case of unobscured ($N_{\rm H} < 10^{22} \rm{cm}^{-2}$) and moderately obscured ($N_{\rm H} = 10^{22} - 10^{23} \rm{cm}^{-2}$) systems. This highlights the importance of further work to better constrain the column density distribution of AGN. Additionally, the eROSITA \citep{Predehl2021} surveys have the potential to address this issue by providing large numbers of moderately obscured AGN out to high redshift and higher $L_X$.   

At higher but still CTN levels of obscuration ($N_{\rm H}=\rm 10^{23}-10^{24}\, \rm cm^{-2}$), our measurements are in fair agreement with previous analytic XLF determinations within the uncertainties of the individual data points. For CTK column densities, our analysis yields only upper limits to the space density. It is nevertheless interesting that for redshifts $z>0.5$ and luminosities $L_X \approx 10^{44} - 10^{45} \, \rm{erg \,s}^{-1}$, the $3\sigma$ upper limits  overlap with the \cite{Ueda_2014} and \cite{Aird_2015} XLFs and therefore provide informative constraints on the space density of AGN with $N_{\rm H}>\rm 10^{24}\, \rm cm^{-2}$. 

An alternative approach for quantifying the CTK AGN demographics is via their fraction relative to the overall AGN population (see Equation \ref{eq_fCT}). Figure \ref{fig:CT_fraction} shows previous estimates of this fraction in comparison  with our constraints. At low redshift, $z<0.5$,  our results favour low CTK fractions, although the uncertainties remain large, $f_{CTK}=21.0^{+16.1}_{-9.9}\%$. 
\cite{Burlon_2011} used the 3-year \textit{Swift}-BAT \citep[Burst Alert Telescope;][]{Barthelmy_2005} serendipitous survey to measure a CTK fraction of about 20\% in the local Universe. Their sample is selected at hard energies (14-195\,keV) and is, therefore, least biased by obscuration effects. 
Low CTK fractions in the range $f_{CTK}\approx10-20\%$ are also proposed by \cite{Akylas_2016} and \cite{Georgantopoulos_2019} based on AGN selected from the 70-month \textit{Swift}-BAT survey. These authors also emphasise the importance of the assumptions on the shape of the CTK AGN X-ray spectrum (e.g. the strength of reflection relative to the direct component, high-energy cut-off) for interpreting the high energy spectra of AGN and deriving CTK fractions. 

\cite{Buchner_2015} analysed the X-ray spectrum of AGN detected in popular extra-galactic survey fields. They constrained the AGN space density using an non-parametric approach, similar to that presented here, but also imposing two different continuity priors. They tend to keep either the value or the slope of the XLF constant in areas of the parameter space with few data to provide meaningful constraints. Figure \ref{fig:CT_fraction} shows the CTK fraction of each of these two priors. These are estimated by marginalising the \cite{Buchner_2015} posteriors in the luminosity interval $10^{42}-10^{46}\rm erg\,s^{-1}$. At low redshift, $z<0.5$, our results are lower but still marginally consistent with the constant-value prior estimates of \cite{Buchner_2015}.  

At higher redshift, $z>0.5$, our $3\sigma$ upper limits also favour the constant-value prior estimates of \cite{Buchner_2015}, at least up to $z\approx2.5$. These upper limits are also broadly consistent with the constraints presented by \cite{Brightman_2012} and also track the $f_{CTK}$ redshift evolution inferred by \citetalias{Lanzuisi_2018}. We cautions that the latter observational constraint corresponds to a luminosity of $L_X(\rm 2-10\,keV) =10^{45}\rm erg\,s^{-1}$ at all redshifts. 
It is also worth noting that the difference between the \cite{Ueda_2014} and \cite{Aird_2015} curves in Figure \ref{fig:CT_fraction} is because of differences in the space density of CTN AGN  ($N_{\rm H}=\rm 10^{22}-10^{24}\, \rm cm^{-2}$). This is also evident by the somewhat higher normalisation of the  \cite{Aird_2015} XLF relative to that \cite{Ueda_2014} in the two middle rows of panels in Figure \ref{fig:space_density}. 

In Figure \ref{fig:CT_fraction}, there is a disagreement between our results and the recent estimates of the CTK fraction of \cite{Ananna2019}. They developed an AGN population synthesis model that fits observations of the diffuse X-ray background spectrum using as input the X-ray luminosity function and models for the X-ray spectra of AGN. Their best-fit model predicts CTK AGN fractions of  $50\pm 9 \%$ at $z\simeq 0.1$, which is higher than our estimates, but also other studies that use spectral analysis of X-ray selected samples to infer CTK fractions directly \citep[e.g.][]{Ricci_2015}.  
We caution that models of the diffuse X-ray background spectrum suffer strong degeneracies between the adopted shape of the X-ray spectra of AGN (i.e. strength of the reflection component, power-law index) and the assumed CTK fraction \citep{Treister2009}.  \cite{Akylas_2012}, for example, showed that low CTK fractions could be compensated by a stronger X-ray reflection component (and vice versa) to yield diffuse X-ray background spectra consistent with observations.

The inferred CTK fractions also have implications on fundamental properties of the accretion flow onto supermassive black holes, such as the radiative efficiency, $\epsilon$. \cite{Shankar_2020} developed models that allow the determination of this parameter based on the observed population properties of AGN samples. They argue in particular that observational measurements of the mean X-ray luminosity averaged over galaxy populations as a function of stellar mass \citep[e.g.][]{Yang2018} provide interesting constraints to $\epsilon$. Their analysis suggests  $\epsilon \gtrsim 0.15$ for the black-hole mass vs stellar mass relation of \cite{Shankar2016}, in line with theoretical expectations. A significant source of uncertainty in this analysis is the fraction of heavily obscured and CTK AGN that may be underrepresented in X-ray-selected samples because of their apparent faintness. If there is a large population of such missing sources, then observational measurements of the mean X-ray luminosity as a function of stellar mass are biased, now leading to the underestimation of the AGN radiative efficiency. Our findings for moderate fractions of Compton thick AGN, at least at $z\approx0.1$, suggest that such an effect is small and hence has a minor effect on the inferred $\epsilon$ values. 

Additionally, within the context of AGN/galaxy co-evolution scenarios, there are suggestions that heavily obscured AGN represent an important early stage of black-hole growth \cite[e.g.][]{Hopkins2008}. They are systems observed close to the peak of their nuclear (and star-formation) activity, at a stage just before the AGN winds blow away the obscuring dust and gas clouds and eventually quench the star-formation in the host galaxy. The low fractions of Compton thick AGN estimated in our work have implications for the duty cycle of the obscured phase of the scenario above. Additionally, our proposed methodology can help isolate reliable samples of heavily obscured and CTK AGN to test the co-evolution scenario above by studying the properties of their host galaxies relative to less obscured sources.

    \subsection{Future missions}\label{Athena}

One of the core science objectives of the \textit{Athena} X-ray observatory \citep{Nandra_2013} is the characterisation of the demographics of heavily obscured and CTK AGN out to high redshift, $z\approx4$ \citep{Georgakakis_2013}. The means to achieve this goal are multi-tiered surveys with the Wide Field Imager \citep[WFI,][]{Rau2013} onboard the \textit{Athena} X-ray observatory. The larger collecting area of \textit{Athena} in the energy range 0.5-10\,keV (2500 cm$^2$ at 6\,keV) compared to e.g. {\it Chandra}, 28 cm$^2$ at 6.5\,keV, or {\it XMM-Newton}, 900 cm$^2$ at 7\,keV, translates into a significant improvement in X-ray spectral quality and ability to identify and characterise the intrinsic properties of heavily obscured and CTK sources. 

\begin{figure}
    \centering
    \includegraphics[width=0.45\textwidth]{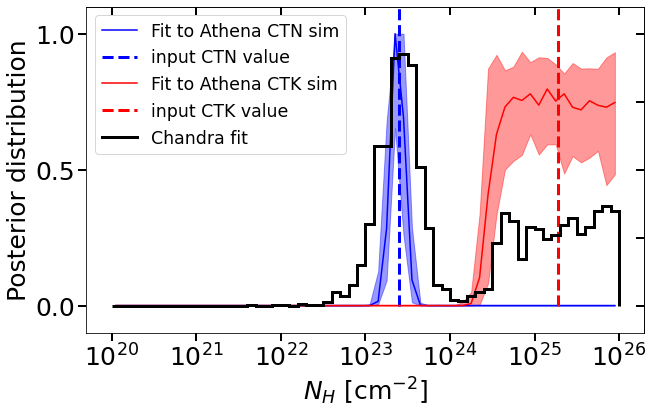}
    \caption{The black histogram shows the  $N_{\rm H}$ posterior distribution of the X-ray spectroscopic fit of \textit{Chandra} COSMOS\_1\_420 without using prior. From the CTN peak (CTK peak, respectively), we simulate an \textit{Athena}/WFI source at $z=1.55$ with ${N_{\rm H}=2.51 \times 10^{23} {\rm cm}^{-2}}$ and ${L_X=1.07 \times 10^{44}{\rm erg\, s}^{-1}}$ (${N_{\rm H}=1.86 \times 10^{25}{\rm cm}^{-2}}$, ${L_X=2.09 \times 10^{45}{\rm erg\, s}^{-1}}$, respectively). The spectroscopic $N_{\rm H}$ posterior distribution of these simulations are represented in blue (red, respectively). The vertical dashed lines correspond to the input $N_{\rm H}$ values for the respective simulations.}
\label{fig:Athena_CTN_CTK}
\end{figure}

We demonstrate this capacity using one of the {\it Chandra} COSMOS Legacy sources with bi-modal column density distribution that includes both CTK and CTN sources. The source ID is COSMOS\_1\_420, with a spectroscopic redshift of 1.55.
Figure \ref{fig:Athena_CTN_CTK} shows the posterior distribution of that source using the  {\it Chandra} COSMOS Legacy observations without applying a $L_{6\mu m}$ prior. 
It also displays the most likely CTK and CTN solutions which are ($N_{\rm H}=1.86 \times 10^{25}$cm$^{-2}$, $L_X=2.09 \times 10^{45}$erg s$^{-1}$) and ($N_{\rm H}=2.51 \times 10^{23}$cm$^{-2}$, $L_X=1.07 \times 10^{44}$erg s$^{-1}$), respectively. We then use the current \textit{Athena} calibration files\footnote{\label{url}https://www.mpe.mpg.de/ATHENA-WFI/response\_matrices.html} to independently simulate the X-ray spectra of both solutions with the UXCLUMPY model. 

In this exercise, we adopt an exposure time of 100\,ks that approximately corresponds to the wide-area tile of the \textit{Athena}/WFI survey plan. The adopted background model is an updated version of the WFI background file generation model\footnote{WFI-MPE-ANA-0010\_i7.1\_Preparation-of-Background-Files.pdf}. 
We assume an extraction aperture of 5\,arcsec that corresponds to an EEF of $\sim 69\%$ for the FOV-average \textit{Athena} PSF. The same model is then used with BXA to refit the simulated spectra. The resulting posteriors are also shown in Figure \ref{fig:Athena_CTN_CTK}. It is clear that for luminous X-ray sources like the COSMOS\_1\_420 example, X-ray spectroscopy with the \textit{Athena}/WFI alone is sufficient to yield unimodal posteriors and distinguish between CTK and CTN solutions. 

\begin{figure}
    \centering
    \includegraphics[width=0.45\textwidth]{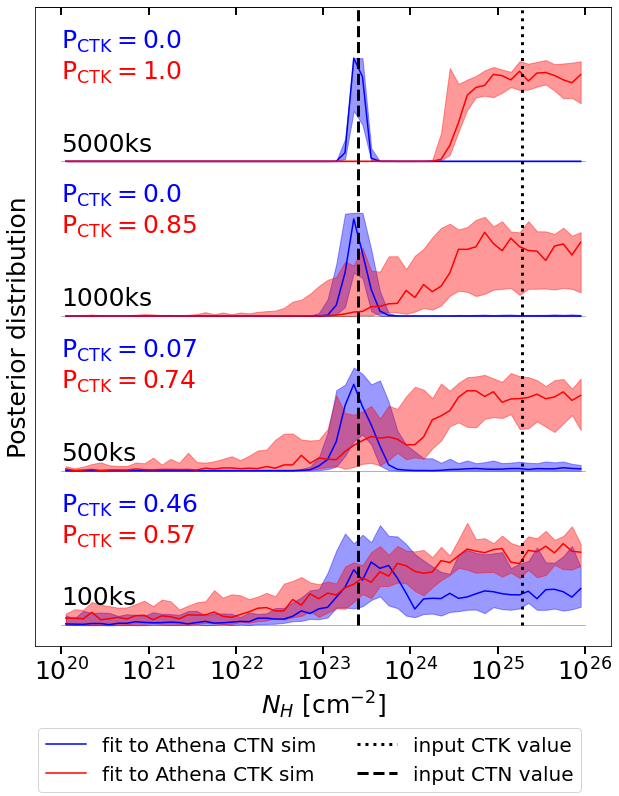}
    \caption{Column density posterior distribution of the X-ray spectroscopic fit for the CTN (blue) and CTK (red) \textit{Athena}/WFI simulations of COSMOS\_1\_420 at $z=1.55$. 
    The intrinsic X-ray luminosity are renormalised at ${L_X(\rm 2-10\,keV)=10^{43}\, erg \, s^{-1}}$ and at ${L_X(\rm 2-10\,keV)=10^{44}\, erg \, s^{-1}}$ for the CTN and CTK solutions, respectively. The distributions are shifted as a function of their exposure time for \textit{Athena}/WFI: 100ks, 500ks, 1000ks and 5000ks. The dashed vertical line represents the CTN input column density at ${N_{\rm H}=2.51 \times 10^{23}}$cm$^{-2}$ and the dotted vertical line represent the CTK input column density at ${N_{\rm H}=1.86 \times 10^{25}}$cm$^{-2}$. P$_{\rm CTK}$, the fraction of the posterior distribution in the CTK regime, is indicated for each exposure time and for both the CTN and CTK \textit{Athena} simulation with their respective colour.}
    \label{fig:Athena_CTK_CTN_renorm}
\end{figure}

Nevertheless, the Athena/WFI surveys will also be sensitive to heavily obscured AGN of lower luminosity. We therefore repeat the analysis by renormalising the X-ray luminosity of the CTN solution of the source COSMOS\_1\_420 to $L_X(\rm 2-10\,keV)=10^{43}\, erg \, s^{-1}$ ($\sim 10$ times fainter) and the CTK solution to $L_X(\rm 2-10\,keV)=10^{44}\, erg \, s^{-1}$ ($\sim 20$ times fainter). The resulting posteriors for different exposure times are shown in Figure \ref{fig:Athena_CTK_CTN_renorm}. 
Short exposure times result in low photon counts leading to broader column density probability distributions that make the determination of the CTK nature of the source uncertain. For instance, with 100ks exposure time, the fits of the simulation from the CTN and CTK solutions have respectively 46\% and 57\% of their posterior distribution in the CTK regime. In this example, the obscuration regime of the source is highly uncertain.

The posterior broadening can be compensated by a  higher exposure time (e.g. $\ga 1000$\,ks for the CTN source or $\ga 5000$\,ks for the CTK source). 
It significantly increases the contrast between the posterior distributions from the 2 simulations. However, such large exposure time are unrealistic for large surveys as planned for \textit{Athena}.
Alternatively, the methodology based on mid-IR priors proposed in this paper could narrow down the posteriors and guarantee the obscuration regime of the source. Doing so would improve the \textit{Athena} constraints on AGN demographics. 

In addition to the \textit{Athena} observatory, observations by the James Webb Space Telescope \citep{Gardner_2006} will improve SED constraints and allow a better understanding of the correlation between X-rays and the mid-IR, on which our methodology hinges. The Euclid survey \citep{Laureijs_2011} will also provide improved photometric redshifts over large areas of the sky, particularly for the obscured AGN population, for which the optical bands are dominated by stellar light of the galaxy.

\section{Summary}\label{Summary_sec}

To understand the growth of SMBH throughout the Universe, one has to get a complete census of AGN and efficiently constrain their physical parameters. This paper proposes a new methodology to extract the properties of X-ray selected AGNs within the \textit{Chandra} COSMOS Legacy survey containing 2965 sources (section \ref{Data_sec}). To our knowledge, this is the first time that these observations have been used to constrain the AGN X-ray luminosity function. The novelty of our analysis is the inclusion of mid-IR priors in the Bayesian-based X-ray spectral analysis (section \ref{L6improve_subsec}). With SED fitting, we constrain the luminosity of the AGNs at $6\mu m$ that is later used as a proxy for the accretion X-ray luminosity guiding the X-ray spectroscopy. This approach improves the confidence of the constraints on physical parameters by breaking down degeneracies, such as between X-ray luminosity and LOS hydrogen column density. This approach primarily benefits the low photon statistics and the most heavily obscured sources. By carefully considering the X-ray selection function, we measure the AGN space density as a function of the accretion X-ray luminosity, LOS obscuration and redshift. We also estimate the CTK fraction as a function of the redshift (section \ref{Obscured_demo_sec}). The main results of our new analysis are:
\begin{itemize}
    \item Our AGN space density measurements are in broad agreement with previous analytic studies. As we find a small number of CTK AGN (27 sources) in the \textit{Chandra} COSMOS Legacy field, we can only place upper limits on the space density of this population. (section \ref{space_density_subsec})
    \item Our CTK fractions estimation are at the low-end of the range determined in previous studies.  At redshift $z < 0.5$,  we find  ${f_{CTK} = 21.0^{+16.1}_{-9.9}\%}$. At $0.5<z<2.5$, we determine (3$\sigma$) upper limits that suggest a CTK fraction typically lower than 40\%, lower than several previous studies. (section \ref{CT_fraction_results_subsec})
    \item By simulating spectra, we found that future missions like the \textit{Athena} observatory would benefit from this multiwavelength methodology to better constrain the physical parameters of the faintest and most obscured sources. (section \ref{Athena})
\end{itemize}

The multiwavelength-based methodology proposed in this paper efficiently increases the confidence of obscuration measurements and AGN demographics. Our results can be used in various fields of SMBH research like their growth, their evolution through time or the AGN co-evolution with the host-galaxy.

\section*{Acknowledgements}

We thank the anonymous referee for their comments and suggestions.
This work has been supported by the EU H2020-MSCA-ITN-2019 Project 860744 “BiD4BESt: Big Data applications for black hole Evolution STudies.”
This research made use of Astropy,\footnote{http://www.astropy.org} a community-developed core Python package for Astronomy \citep{astropy:2013, astropy:2018}. 
This research has made use of data obtained from the Chandra Data Archive and the Chandra Source Catalog, and software provided by the Chandra X-ray Center (CXC) in the application packages CIAO and Sherpa.
For analysing X-ray spectra, we use the analysis software BXA \citep{Buchner_2014}, which connects the nested sampling algorithm UltraNest \citep{Buchner_2021} with the fitting environment CIAO/Sherpa \citep{Fruscione2006}.

DMA and DJR acknowledges the Science Technology and Facilities Council (STFC) for support through grant codes ST/T000244/1.

JA acknowledges support from a UKRI Future Leaders Fellowship (grant code: MR/T020989/1).

FJC acknowledges financial support from the Spanish Ministry MCIU under project RTI2018-096686-B-C21 (MCIU/AEI/FEDER/UE), cofunded by FEDER funds and from the Agencia Estatal de Investigación, Unidad de Excelencia María de Maeztu, ref. MDM-2017-0765.

AL is partly supported by the PRIN MIUR 2017 prot. 20173ML3WW 002 ‘Opening the ALMA window on the cosmic evolution of gas, stars, and massive black holes’.

CRA acknowledges support from the projects ``Feeding and feedback in active galaxies'', with reference
PID2019-106027GB-C42, funded by MICINN-AEI/10.13039/501100011033, ``Quantifying the impact of quasar feedback on galaxy evolution'', with reference EUR2020-112266, funded by MICINN-AEI/10.13039/501100011033 and the European Union NextGenerationEU/PRTR, and from the Consejer\' ia de Econom\' ia, Conocimiento y Empleo del Gobierno de
Canarias and the European Regional Development Fund (ERDF) under grant ``Quasar feedback and molecular gas reservoirs'', with reference ProID2020010105, ACCISI/FEDER, UE.

For the purpose of open access, the authors have applied a Creative Commons Attribution (CC BY) licence to any Author Accepted Manuscript version arising from this submission.

\section*{Data Availability}

We release two tables containing information on the sources of our sample. 
The first one contains the information relative to the spectral extraction and is partially displayed in Table \ref{tab:spectral_extraction_info}. The second one contains the results of the fit by BXA and is partially displayed in Table \ref{tab:fit_information}.
We also release the STAN products for the space density and CTK fraction calculations. The files are described and available at \url{https://doi.org/10.5281/zenodo.7014625}.

All other data products used in this paper are available on-demand to the authors.



\bibliographystyle{mnras}
\bibliography{mnras_template} 




\appendix

\bsp	
\label{lastpage}

\end{document}